\documentclass[12pt,letterpaper]{article}
\pdfoutput=1

\usepackage{xcolor}
\usepackage{amsfonts,amssymb,amsmath}
\usepackage{MnSymbol,wasysym}
\usepackage{esint}

\usepackage{graphicx}
\usepackage{amsmath,systeme}
\usepackage{caption}
\usepackage{subcaption}
\usepackage{booktabs}

\usepackage{listings}

\usepackage{hyperref}

\usepackage[style=numeric-comp, sorting=none]{biblatex}
\DeclareFieldFormat{doi}{}
\DeclareFieldFormat{url}{}
\DeclareFieldFormat{eprint}{}

\numberwithin{equation}{section}

\newcommand{\be}{\begin{equation}}
\newcommand{\ee}{\end{equation}}
\newcommand{\ben}{\begin{displaymath}}
\newcommand{\een}{\end{displaymath}}
\newcommand{\bea}{\begin{eqnarray}}
\newcommand{\eea}{\end{eqnarray}}
\newcommand{\bean}{\begin{eqnarray*}}
\newcommand{\eean}{\end{eqnarray*}}

\newcommand{\commentout}[1]{}

\usepackage{amsmath}

\renewcommand{\theequation}{\arabic{section}.\arabic{equation}}

\newcommand{\beq}{\begin{equation}}
\newcommand{\eeq}{\end{equation}}
\newcommand{\beqr}{\begin{displaymath}}
\newcommand{\eeqr}{\end{displaymath}}
\newcommand{\beqa}{\begin{eqnarray}}
\newcommand{\eeqa}{\end{eqnarray}}
\newcommand{\beqar}{\begin{eqnarray*}}
\newcommand{\eeqar}{\end{eqnarray*}}
\newcommand{\mk}[1]{{\color{red} \bf #1}}
\newcommand{\cmmnt}[1]{}

\usepackage{setspace}

\usepackage{float}

\allowdisplaybreaks

\newcounter{customfootnote}  % Create a new counter
\newcommand{\customfootnote}[1]
{%
  \refstepcounter{customfootnote}%
  \footnote[\value{customfootnote}]{#1}%
}

\begin{document}
%%%%%%%%%%%%%%%%%%%%%%%%%%%%%%%%%%%%%%%%%%%%%%%%%%%%%%%%%%%%%%%%%%%%%%%%
%%%%%%%%%%%%%%%%%%%%%%%%%%%%%%%%%%%%%%%%%%%%%%%%%%%%%%%%%%%%%%%%%%%%%%%%
%%%%%%%%%%%%%%%%%%%%%% TITLEPAGE %%%%%%%%%%%%%%%%%%%%%%%%%%%%%%%%%%%%%%%
%%%%%%%%%%%%%%%%%%%%%%%%%%%%%%%%%%%%%%%%%%%%%%%%%%%%%%%%%%%%%%%%%%%%%%%%
%%%%%%%%%%%%%%%%%%%%%%%%%%%%%%%%%%%%%%%%%%%%%%%%%%%%%%%%%%%%%%%%%%%%%%%%
%\onehalfspacing

%\title{\LARGE \bf Stationary and Self Similar solutions of Gross-Pitaevskii Equation in 2D and 3D, and their Borel analysis}
\title{\LARGE \bf Stationary acoustic black hole solutions in Bose-Einstein condensates and their Borel analysis}

\author{
	 Sachin Vaidya$^1$,
	 Martin Kruczenski$^1$\thanks{E-mail: \texttt{vaidya2@purdue.edu, markru@purdue.edu.}} \\
	$^1$ Dep. of Physics and Astronomy, \\ Purdue University, W. Lafayette, IN  \\
	}

\date{}

\maketitle

\begin{abstract}
 In this article, we study the dynamics of a Bose-Einstein condensate (BEC) with the idea of finding solutions that could possibly correspond to a so-called acoustic (or Unruh) black/white holes. Those are flows with horizons where the speed of the flow goes from sub-sonic to super-sonic. This is because sound cannot go back from the supersonic to the subsonic region. The speed of sound plays the role of the speed of light in a gravitational black hole, an important difference being that there are excitations that can go faster than the speed of sound and therefore can escape the sonic black hole. Here, the motion of the BEC is described by the Gross-Pitaevskii Equation (GPE). More concretely, we discuss singular Stationary solutions of Gross-Pitaevskii equation in 2D (with Circular symmetry) and 3D (with Spherical symmetry). We use these solutions to study the local speed of sound and magnitude of flow velocity of the condensate to see whether they cross, indicating the potential existence of a sonic analog of a black/white hole. We discuss numerical techniques used and also study the semi-analytical Laplace-Borel resummation of asymptotic series solutions to see how well they agree with numerical solutions. We also study how the resurgent transseries plays a role in these solutions. 
\end{abstract}

\clearpage
\newpage

%\makeindex
\tableofcontents
%\keywords{Classical string solutions, AdS/CFT, Wilson loops}

%\preprint{\tt{} \\
%          \tt{hep-th/yymmnnn}  }

\onehalfspacing
%%%% INTRODUCTION
%\setcounter{section}{1}
\section{Introduction}
\label{intro}
%\subsection{Gross Pitaevskii Equation}

The main purpose of this paper is to study solutions to a differential equation that appears in the context of black hole simulations using superfluids. In \cite{PhysRevLett.46.1351} Unruh proposed the idea of studying acoustic black holes as a way to bring quantum phenomena such as Hawking radiation \cite{HawkingS.W.1975Pcbb} closer to experimentation. This has been extensively studied theoretically in the context of various analog gravity studies (see \cite{Barceló2011, CarlosBarceló_2001, Visser2002, PhysRevLett.85.4643, PhysRevD.105.124066, Tian2022}) as well as in analog cosmological models (see \cite{PhysRevA.69.033602}, \cite{PhysRevLett.91.240407}). This challenge was subsequently taken up by various experimental groups, for example \cite{Steinhauer2014} and \cite{MunozdeNova2019}, and more recently \cite{Tamura:2023mby}. A setup where one expects to see such quantum phenomena is in acoustic black holes produced by the motion of a superfluid (Bose-Einstein condensate) at ultra-cold temperatures. In this work, we consider the Gross-Pitaevskii equation (GPE) with coupling $g$ and without an external potential \cite{NozieresPinesBEC}, that describes the dynamics of a Bose-Einstein condensate (BEC) 
\begin{equation} \label{e:GPE} 
%\begin{align}
i \hbar  \frac{\partial \psi  \left(\Vec{\mathbf{r}} , t\right)}{\partial t} = -\frac{\hbar^{2}}{2 m} 
\nabla^{2}\psi  \left(\Vec{\mathbf{r}} , t\right) + g {\lvert \psi  \left(\Vec{\mathbf{r}} , t\right)\rvert}^{2} \psi  \left(\Vec{\mathbf{r}} , t\right)
%\end{align}
\end{equation}

Small perturbations follow a dispersion relation \cite{NozieresPinesSound}:

\begin{equation} \label{e:GPE local speed of sound dispersion} 
%\begin{align}
E = \hbar\omega = \sqrt{\frac{\hbar^2 \lvert\Vec{\mathbf{q}}\rvert^2}{2 m} \left(\frac{\hbar^2 \lvert\Vec{\mathbf{q}}\rvert^2}{2 m} + 2gn \right)}
%\end{align}
\end{equation}
where,  $n$ is the density,  $\Vec{\mathbf{q}}$ is momentum, and $\omega$ is the frequency.

From equation \eqref{e:GPE local speed of sound dispersion}, we get $E \approx c \hbar \lvert\Vec{\mathbf{q}}\rvert$  when $ \frac{\hbar^2 \lvert\Vec{\mathbf{q}}\rvert^2}{2 m} << 2gn $ which is a long-wavelength approximation. In that case, the excitations are phonons with speed of sound
\begin{equation} \label{e:speed of sound}
c=\sqrt{\frac{ng}{m}}
\end{equation}

Now if we write the current density 
\beq
\Vec{\mathbf{j}}\left(\Vec{\mathbf{r}} , t \right)=-\frac{i \hbar}{2 m} \left({\psi}^{*}  \left(\Vec{\mathbf{r}} , t\right) \Vec{\nabla}\psi  \left(\Vec{\mathbf{r}} , t\right) - \psi  \left(\Vec{\mathbf{r}} , t\right) \Vec{\nabla}{\psi}^{*}  \left(\Vec{\mathbf{r}} , t\right) \right)
\eeq
from \eqref{e:GPE} and substitute $\psi \left(\Vec{\mathbf{r}} , t\right) = \sqrt{n \left(\Vec{\mathbf{r}} , t\right)}  {\mathrm e}^{i \theta  \left(\Vec{\mathbf{r}} , t\right)}  {\mathrm e}^{-i \mu t}$ into it as in \cite{NozieresPinesVelocity}, we get the fluid velocity as
\begin{equation} \label{e:flow velocity}
\Vec{v} \left(\Vec{\mathbf{r}} , t \right) = \frac{\Vec{\mathbf{j}} \left(\Vec{\mathbf{r}} , t\right)}{n \left(\Vec{\mathbf{r}} , t\right)} = \frac{\hbar}{m} \Vec{\nabla} \theta  \left(\Vec{\mathbf{r}} , t\right)
\end{equation}

Our main focus is to find stationary solutions to the equation \eqref{e:GPE} corresponding to acoustic black holes in two (and three) spatial dimensions. We assume an infinite-size fluid that undergoes an inward radial flow with a magnitude of radial velocity $v(r)$. The magnitude of the inward velocity vanishes at infinity and increases as $r$ decreases. If, below a certain radius $r_h$, the magnitude of inward velocity is larger than the speed of sound \Big($v(r<r_h)>c$\Big) then $r_h$ acts as an acoustic horizon \cite{PhysRevA.99.023850}, and sound waves cannot escape from that region. In principle, the flow can again become subsonic at even smaller radius. In addition, the density will become singular ($\rho\rightarrow\infty$) at some point $r=r_0\geq 0$. In practice, when the density is large, the fluid is removed by atomic collisions so that $\rho$ remains finite. A similar problem is discussed in \cite{PhysRevA.109.023305}, wherein the singularity is removed by including a three-body recombination term and an external potential. However, our study focuses on singular solutions as an approximation to non-singular ones.

We do not expect to find analytical solutions and therefore we employ a combination of asymptotic series expansion and numerical techniques. When the series expansion alone does not contain all the information about the system, more advanced mathematical techniques are required to extract the information about non-perturbative corrections. These non-perturbative corrections show up when the asymptotic perturbative series has factorially divergent coefficients. Furthermore, when the differential equation is nonlinear in the dependent variable, non-perturbative corrections keep appearing with increasing powers and also bring in their own factorially divergent perturbative series multiplying them. This is known as a transseries (\cite{10.1155/S1073792895000286}, \cite{10.1215/S0012-7094-98-09311-5}) that needs to be resummed using the Laplace-Borel transform to get the most general solution. The phenomenon of resurgence in the transseries is also required for the most general solution (\cite{Aniceto2015}, \cite{DORIGONI2019167914}, \cite{Costin_2019}, and \cite{PhysRevD.92.125011}) to be real. This allows us to check properties of the numerical solutions that can be obtained with standard methods such as Newton iteration with Chebyshev collocation.  

Resurgent transseries and its Borel summation have been used for various problems such as the nonlinear equation for Bjorken flow in conformal hydrodynamics \cite{PhysRevD.92.125011}, Painlev\'e equation \cite{Costin_2019} and many others.

\section{Stationary states: acoustic black holes} 
To get the stationary solutions, we start by considering solutions with only radial and time dependence of the form
\beq
\psi  \left(\Vec{\mathbf{r}} , t\right) = {\mathrm e}^{-i \mu  t} \phi  (r)
\eeq
Substituting this into \eqref{e:GPE} we get
\begin{equation} \label{e:GPE2Scaled}
%\begin{align}
\Phi  (R) = -\nabla_R^{2}\Phi  (R)+{\vert \Phi  (R)\rvert}^{2} \Phi  (R)
%\end{align}
\end{equation}

\noindent where the variables were rescaled as, 
\beqa
R&=&\sqrt{\frac{2m\mu}{\hbar}}r, \\
\Phi  (R) &=& \rho(R) e^{i \theta(R)}, \\
\rho  (R) &=& \sqrt{\frac{g}{\mu  \hbar}}\ \sqrt{n  (R)} 
\eeqa
and 
\beq
\nabla_R^{2} \Phi (R) = \frac{1}{R^{\mathbf{d}-1}} \frac{\partial}{\partial R} \left(R^{\mathbf{d}-1} \frac{\partial \Phi (R)}{\partial R} \right)
\eeq
is the radial Laplacian in $\mathbf{d}$ dimensions. Having a homogeneous solution ($\Phi$ independent of $R$) requires $\Phi=0$ (no fluid) or $|\Phi|=1$ as $R\to\infty$. We are interested in solutions that asymptotically at large $R$ describe a static homogeneous fluid and therefore we impose the boundary condition 
\beq
 |\Phi(R\rightarrow\infty)|=1
 \label{rhobc}
\eeq
in all the solutions we consider. 
The local speed of sound (from \eqref{e:speed of sound}) is given by $c = \sqrt{\frac{n  (R)g}{m}}$ \cite{NozieresPinesSound}, which becomes $c = \sqrt{\frac{\hbar\mu}{m}}\rho  (R)$ in the coordinates $R$ and $T$. It is convenient to introduce a rescaled local speed of sound 
\beq \label{e:scaled_sound_speed}
C = \sqrt{\frac{2m}{\hbar\mu}}c = \sqrt{2} \rho  (R)
\eeq
The radial flow velocity (from \eqref{e:flow velocity}) of the Bose-Einstein condensate (BEC) is given by $\Vec{\mathbf{v}}  (r) = \frac{\hbar}{m}\Vec{\nabla} \theta  (r)  = \frac{\hbar}{m}\frac{\partial \theta  (r)}{\partial r} \hat{\mathbf{r}}$   \cite{NozieresPinesVelocity}, which simplifies to $\Vec{\mathbf{v}}  (R) = \sqrt{\frac{2\mu\hbar}{m}}\frac{\partial \theta  (R)}{\partial R} \hat{\mathbf{R}}$ in the coordinates $R$ and $T$. This gives a rescaled radial velocity:
\beq \label{e:scaled_vel}
\Vec{\mathbf{V}}  (R) = \Vec{\mathbf{v}}  (R)\sqrt{\frac{2m}{\mu\hbar}} = 2\frac{\partial \theta  (R)}{\partial R} \hat{\mathbf{R}}
\eeq
For a stationary solution with only radial dependence in $\mathbf{d}$ dimensions, substituting $\Phi  (R) = \rho  (R) \exp(i \theta(R))$ into \eqref{e:GPE2Scaled}, the imaginary part of the equation becomes 
\begin{equation} \label{e:StationaryImGPE2}
%\begin{align}
\left(2 \frac{d \rho}{d R}+\frac{(\mathbf{d}-1) \rho }{R}\right) \left(\frac{d \theta  }{d R}\right)+\rho   \left(\frac{d^{2} \theta  }{d R^{2}}\right)=0
%\end{align}
\end{equation}
and the real part becomes
\begin{equation} \label{e:StationaryReGPE2}
%\begin{align}
\frac{d^{2} \rho }{d R^{2}}+\frac{\left(\mathbf{d}-1\right)}{R} \frac{d \rho  }{d R}+\rho  \left(R \right)-\left(\frac{d \theta  }{d R}\right)^{2} \rho -\rho  ^{3}=0
%\end{align}
\end{equation}
Integrating \eqref{e:StationaryImGPE2} and introducing the constant of integration $B$ gives: 
\beq \label{e:dtheta_dR}
\frac{d \theta  }{d R} = \frac{B}{R^{\mathbf{d}-1} \rho ^{2}}
\eeq
which is proportional to the flow velocity as shown in \eqref{e:scaled_vel}. Therefore, the solutions with $B<0$ have a radially inward velocity (black hole) and the solutions with $B>0$ have a radially outward velocity (white hole).

From equations~\eqref{e:dtheta_dR} and~\eqref{e:StationaryReGPE2}, we now get
\begin{equation} \label{e:StationaryODE}
%\begin{align}
\frac{d^{2} \rho  }{d R^{2}}+\frac{(\mathbf{d}-1)}{R} \frac{d \rho  }{d R}+\rho -\frac{B^{2}}{\rho  ^{3}\, R^{2(\mathbf{d}-1)}}-\rho  ^{3} = 0
%\end{align}
\end{equation}
which is the main equation we have to solve. 
It is interesting to note that equation \eqref{e:StationaryODE} has the same solution for $B<0$ (inward flow) and $B>0$ (outward flow) for given boundary conditions including the condition  $\rho(R\to\infty)\to 1$ discussed in \eqref{rhobc}.

\section{Numerical Solutions}

 We start the discussion of the numerical methods by considering the possible  singularities of the solutions. A short analysis of \eqref{e:StationaryODE}, shows that in 2D and 3D,  a singularity can appear at a finite value $R=R_0>0$:
\begin{equation} \label{e:sqrt(2)/(R-R0)}
\frac{d^{2} \rho \left(R \right)}{d R^{2}}-\rho \left(R \right)^{3} \approx 0 \implies \rho \left(R \right) \approx \frac{\sqrt{2}}{R-R_{0}} \hspace{0.5cm} \text{ when, $\mathbf{d} = 2$, $3$}
\end{equation}
 For $\mathbf{d}=2$ it is also possible to have a singularity at the origin $R=0$ of two types:
\begin{equation} \label{e:1/R}
\frac{d^{2} \rho \left(R \right)}{d R^{2}}+\frac{(\mathbf{d}-1)}{R} \frac{d \rho \left(R \right)}{d R}-\rho \left(R \right)^{3} \approx 0 \implies \rho \left(R \right) \approx \frac{1}{R} 
\end{equation}
\begin{equation} \label{e:ln(R)}
\frac{d^{2} \rho \left(R \right)}{d R^{2}}+\frac{(\mathbf{d}-1)}{R} \frac{d \rho \left(R \right)}{d R}-\frac{B^{2}}{\rho \left(R \right)^{3} R^{2(\mathbf{d}-1)}} \approx 0 
\\  \implies \rho \left(R \right) \approx C_{0} \ln \! \left(R \right)
\end{equation}
where  $C_{0}$ is an arbitrary constant. 

For the actual numerical work, it is convenient to use the function $f(R)=\frac{1}{\rho(R)}$ which vanishes at the singular point. Also, it does not diverge anywhere since, for physical solutions,  the density is never zero ($\rho(R)>0$). It is also convenient to map the $R$ coordinate to a coordinate $U$ defined through
\beq
R = R_0 + A \frac{1+U}{1-U}
\eeq
such that $-1\le U \le 1$. Here $R_0\ge0$ is the arbitrarily chosen location of the singularity that appears as a new integration constant of the solutions. As shown later, there is a range of allowed values of $R_0$ that depends on the constant $B$. The parameter $A$ that we take here to be $A=10$ appears for numerical reasons.

\begin{figure}[H]
    \centering
    \begin{subfigure}{1\textwidth}
        \centering
        \includegraphics[width=\linewidth]{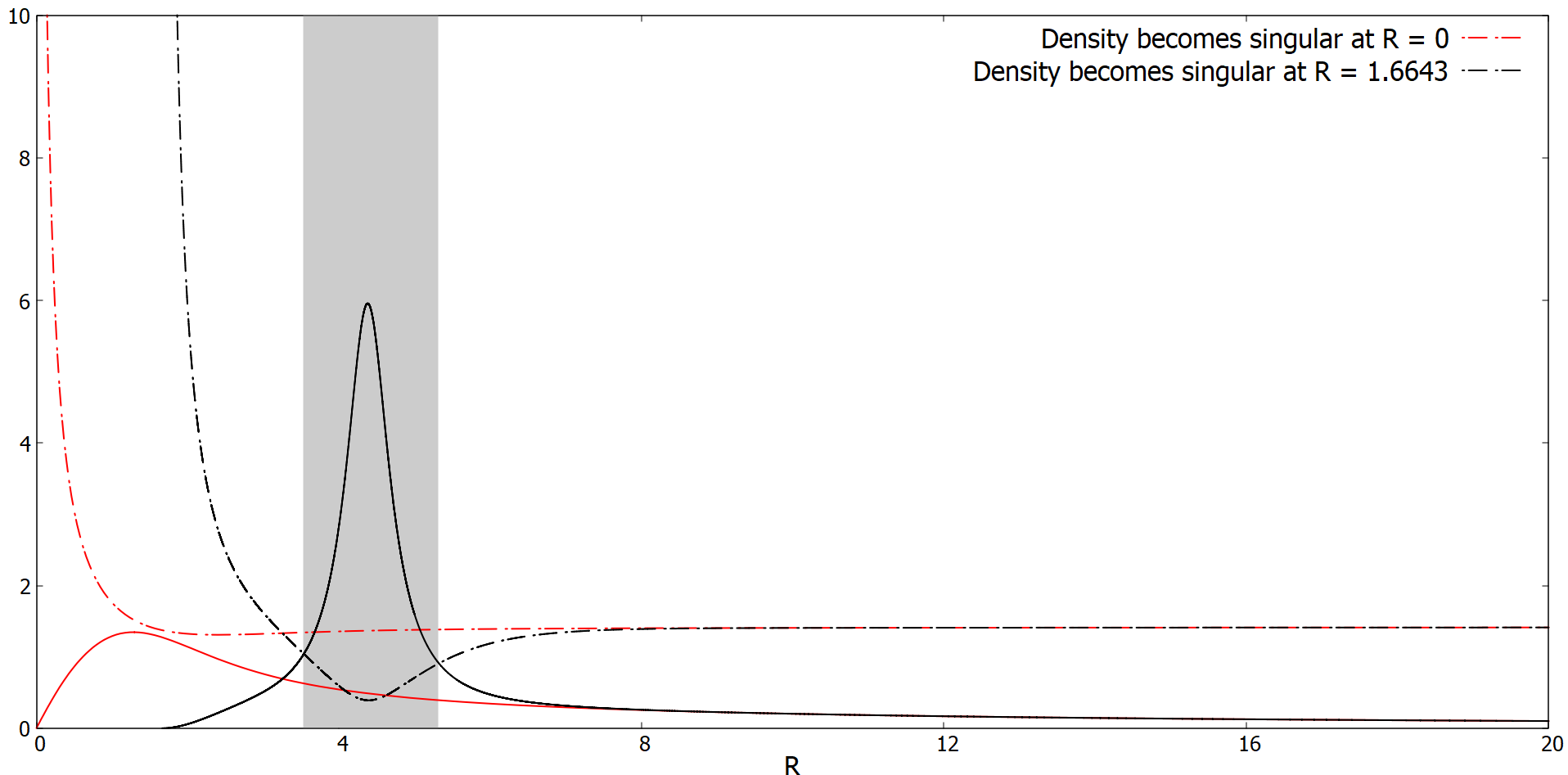}
        \caption{$\lvert B\rvert = 1$ in 2D} \label{fig:B1_2D}
    \end{subfigure}
    \\[10pt]
    \begin{subfigure}{1\textwidth}
        \centering
        \includegraphics[width=\linewidth]{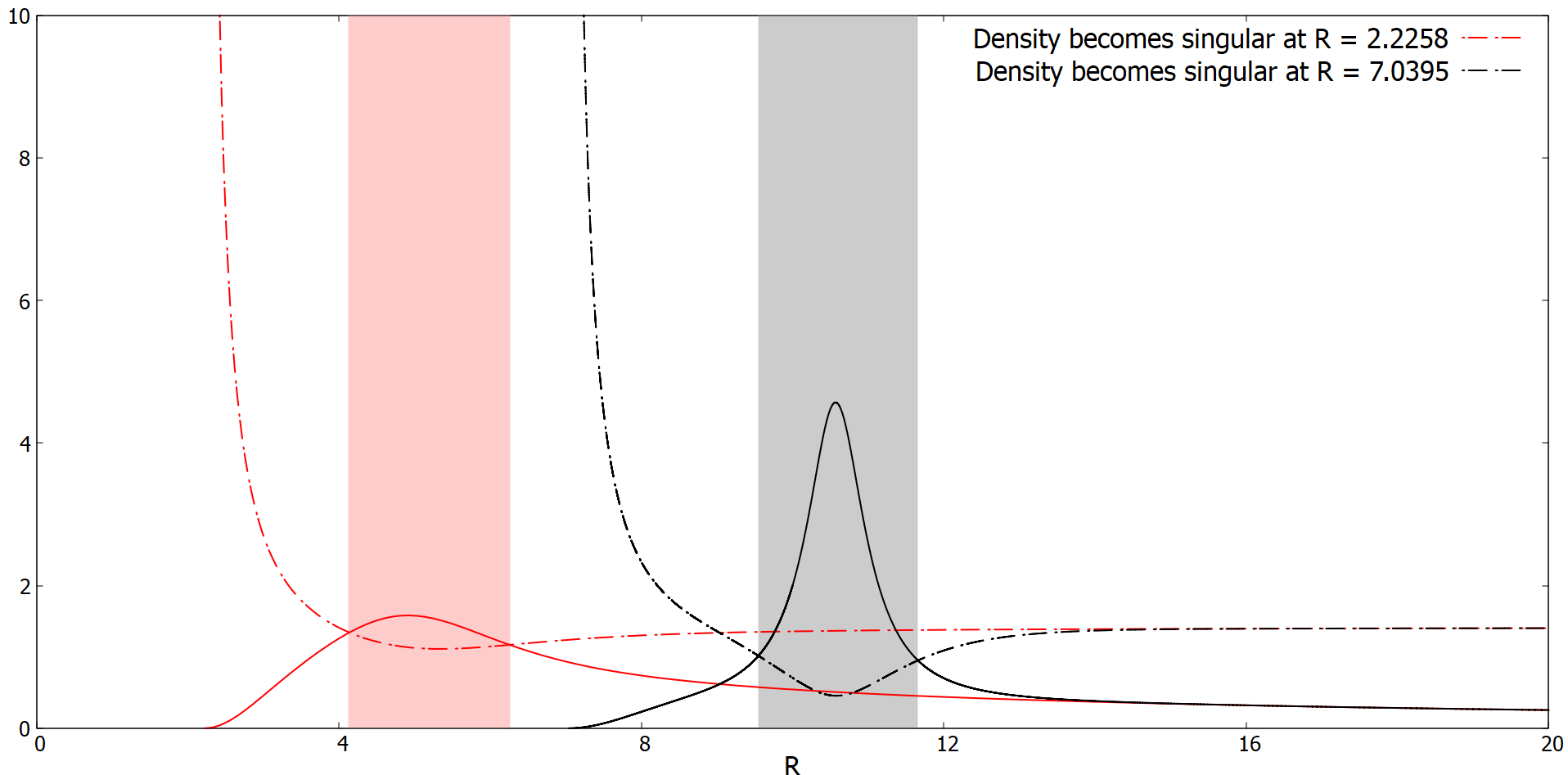}
        \caption{$\lvert B\rvert= 2.5$ in 2D} \label{fig:B2.5_2D}
    \end{subfigure}
    \vspace{5pt}
    \caption{Scaled local speed of sound (dashed line) and magnitude of flow velocity (solid line) of BEC in 2D, obtained by solving $\mathbf{1)}$ BVP using newton iteration (in red), $\mathbf{2)}$ BVP+IVP using Newton iteration and RK4 (in black). Shaded regions are supersonic regions for respective solutions.}
    \label{fig:2D_numerical}
\end{figure}
\begin{figure}[H]
    \centering
    \begin{subfigure}{1\textwidth}
        \centering
        \includegraphics[width=\linewidth]{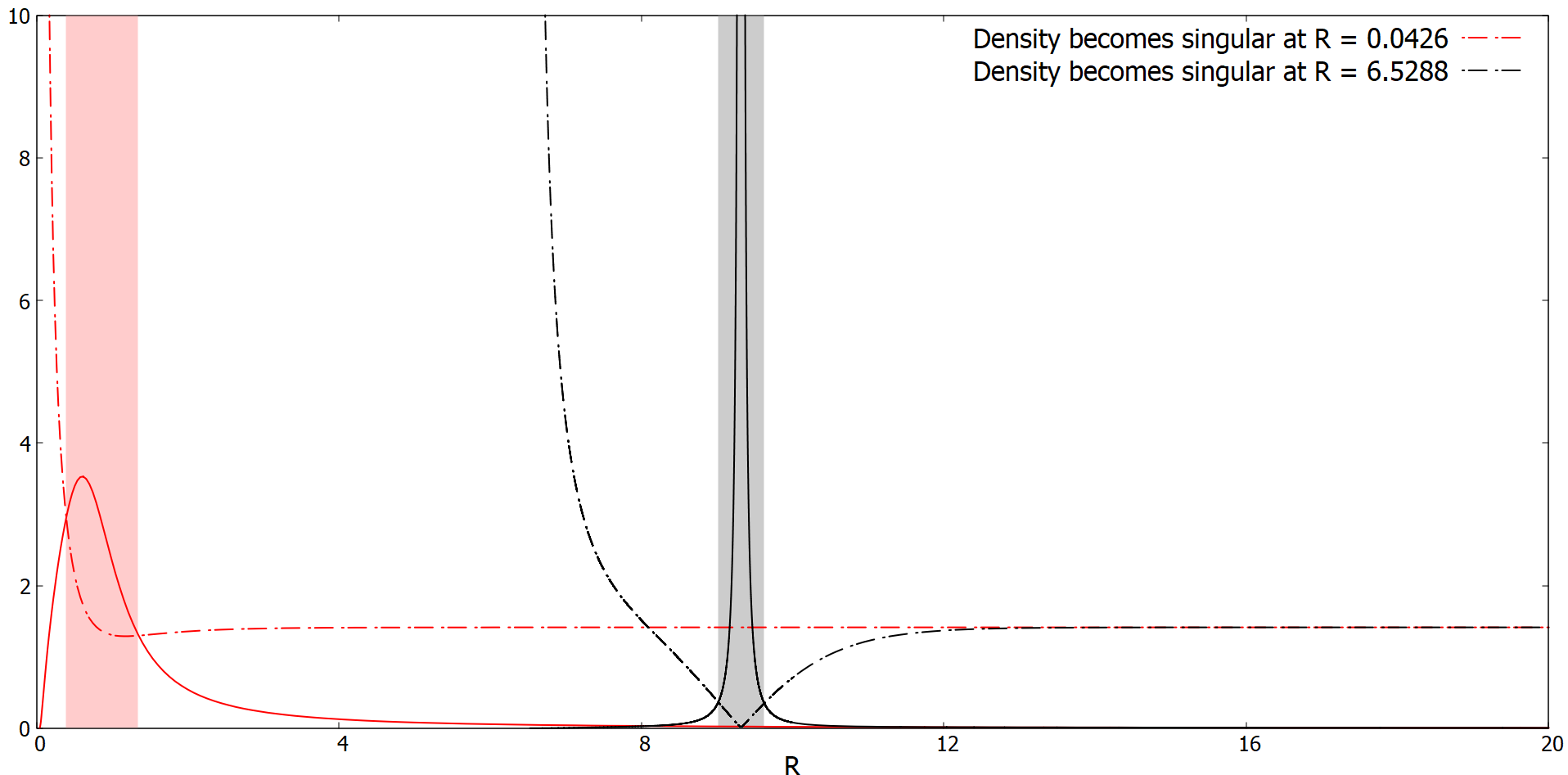}
        \caption{$\lvert B\rvert= 1$ in 3D} \label{fig:B1_3D}
    \end{subfigure}
    \\[10pt]
    \begin{subfigure}{1\textwidth}
        \centering
        \includegraphics[width=\linewidth]{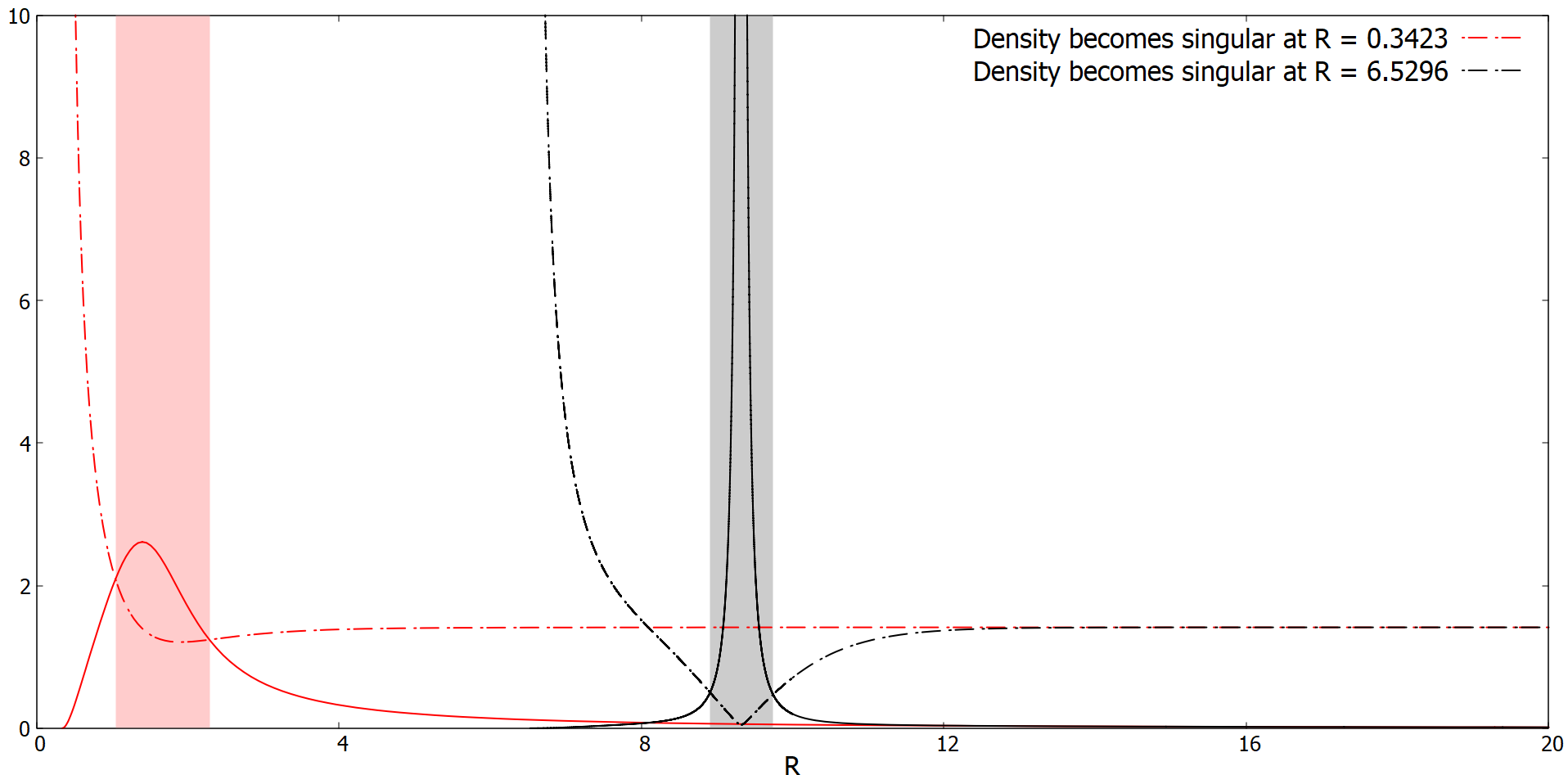}
        \caption{$\lvert B\rvert = 2.5$ in 3D} \label{fig:B2.5_3D}
    \end{subfigure}
    \vspace{5pt}
    \caption{Scaled local speed of sound (dashed line) and magnitude of flow velocity (solid line) of BEC in 3D, obtained by solving $\mathbf{1)}$ BVP using newton iteration (in red), $\mathbf{2)}$ BVP+IVP using newton iteration and RK4 (in black). Shaded regions are supersonic regions for respective solutions.}
    \label{fig:3D_numerical}
\end{figure}

\begin{figure}[H]
    \centering
    \begin{subfigure}{1\textwidth}
        \centering
        \includegraphics[width=\linewidth]{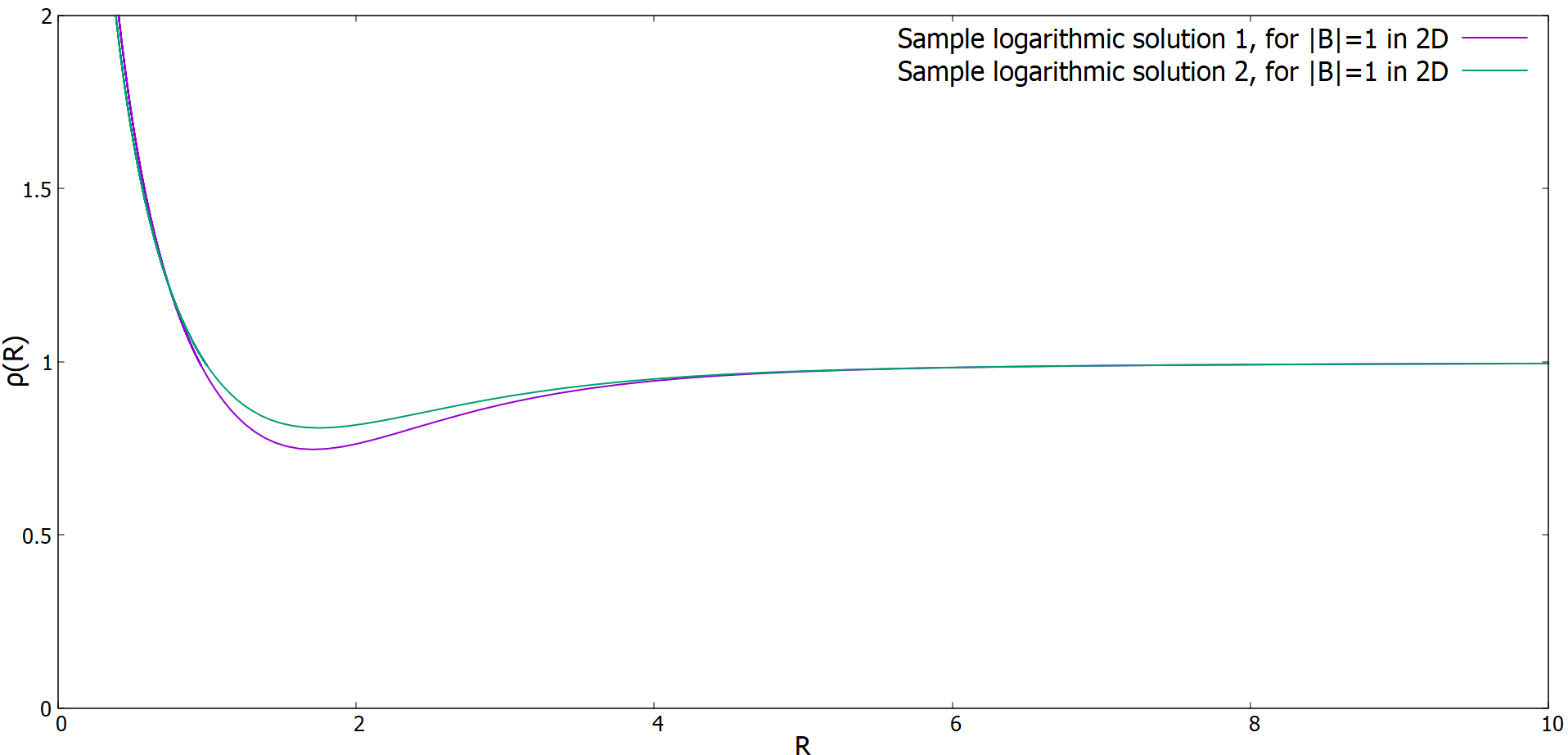}
        \caption{Sample logarithmic solutions} \label{fig:ln_a}
    \end{subfigure}
    \\[10pt]
    \begin{subfigure}{1\textwidth}
        \centering
        \includegraphics[width=\linewidth]{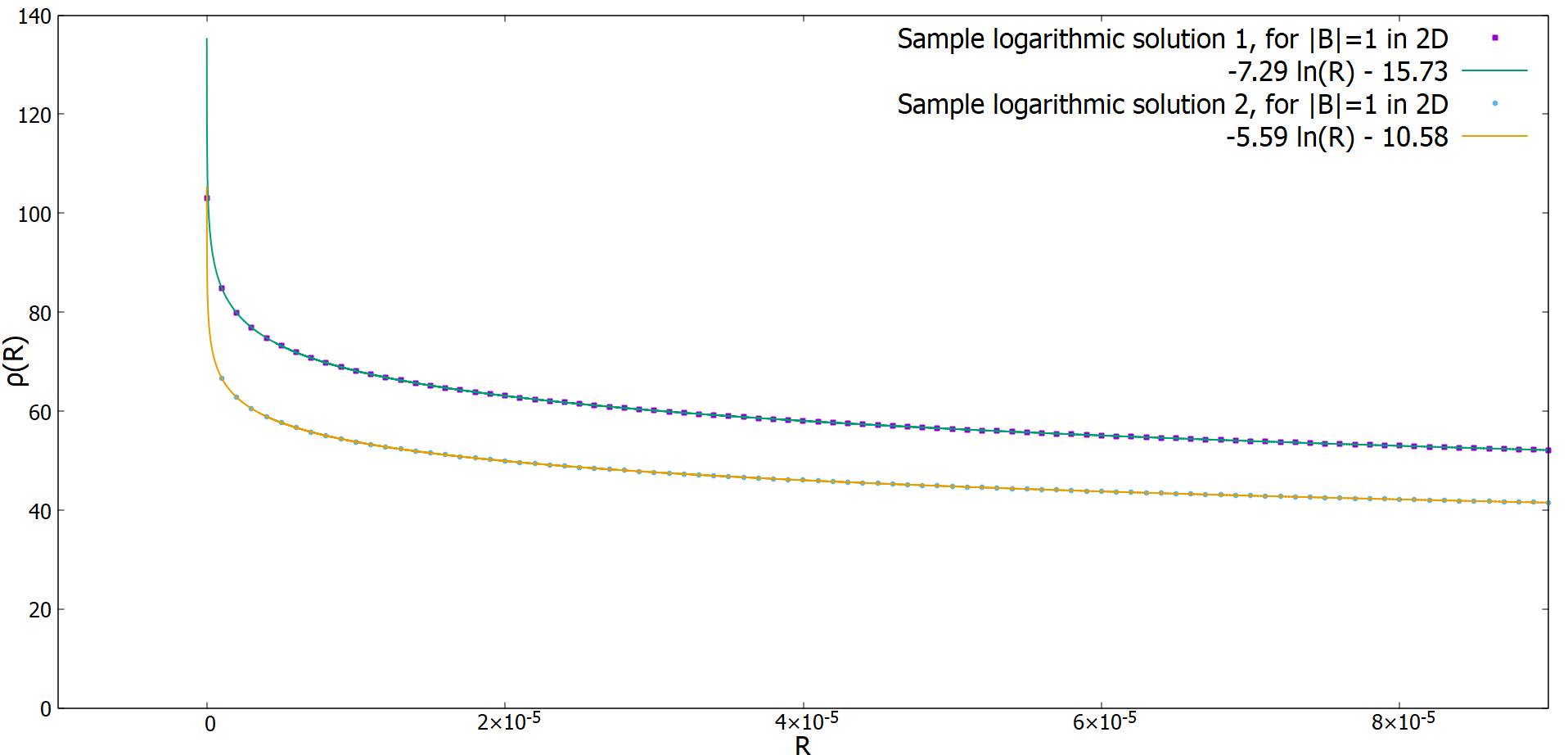}
        \caption{Zoomed in near the singularity comparing the numerical solution (dots) with the predicted asymptotic logarithmic behavior using a fit} \label{fig:ln_b}
    \end{subfigure}
    \vspace{5pt}
    \caption{Sample solutions with logarithmic singularities at $R=0$ in 2D, for $\lvert B\rvert=1$}
    \label{fig:2D_numerical_ln}
\end{figure}
\begin{figure}[H]
    \centering
    \begin{subfigure}{1\textwidth}
        \centering
        \includegraphics[width=\linewidth]{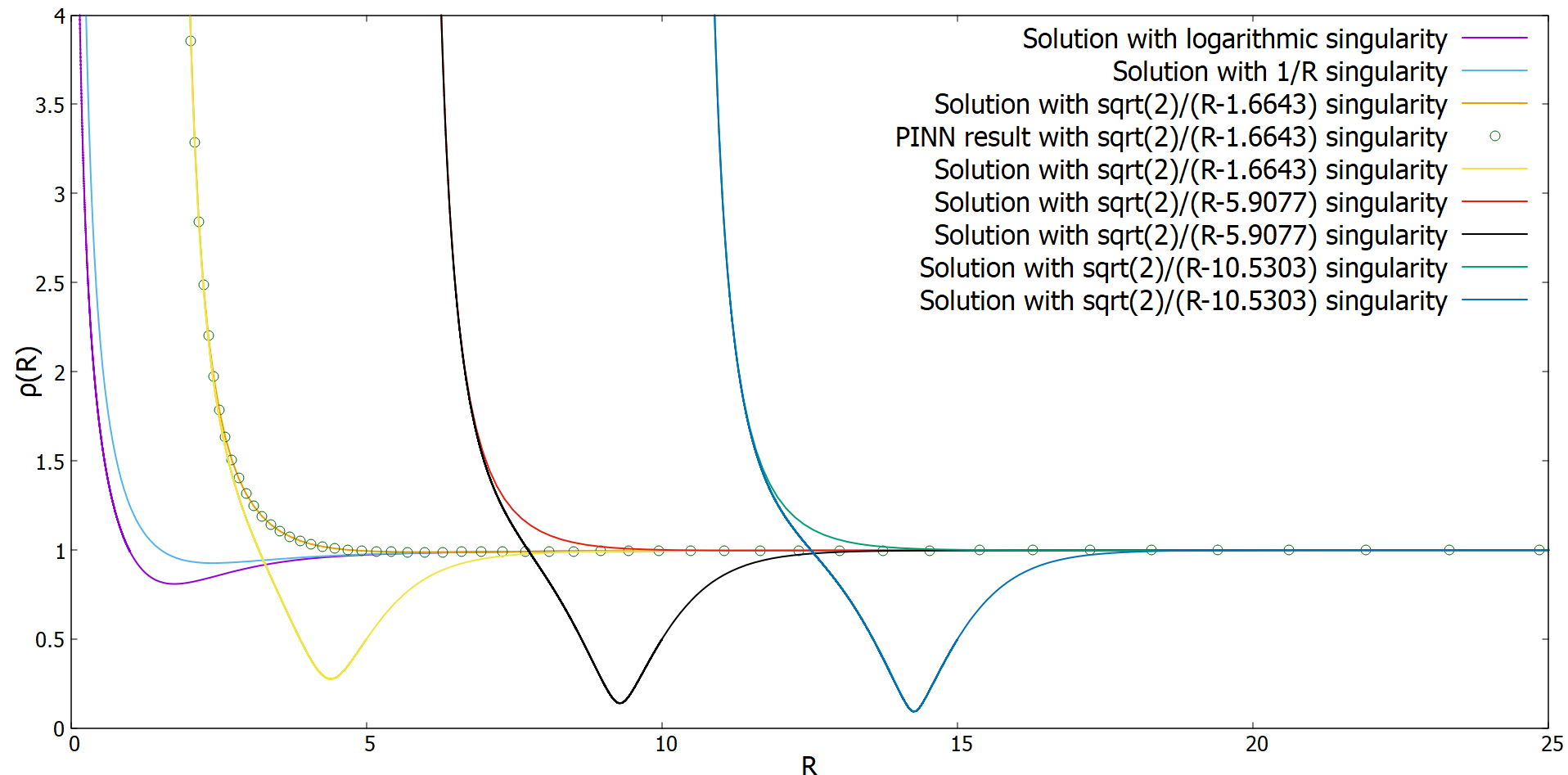}
        \caption{In 2D} \label{fig:nonunique_a}
    \end{subfigure}
    \\[10pt]
    \begin{subfigure}{1\textwidth}
        \centering
        \includegraphics[width=\linewidth]{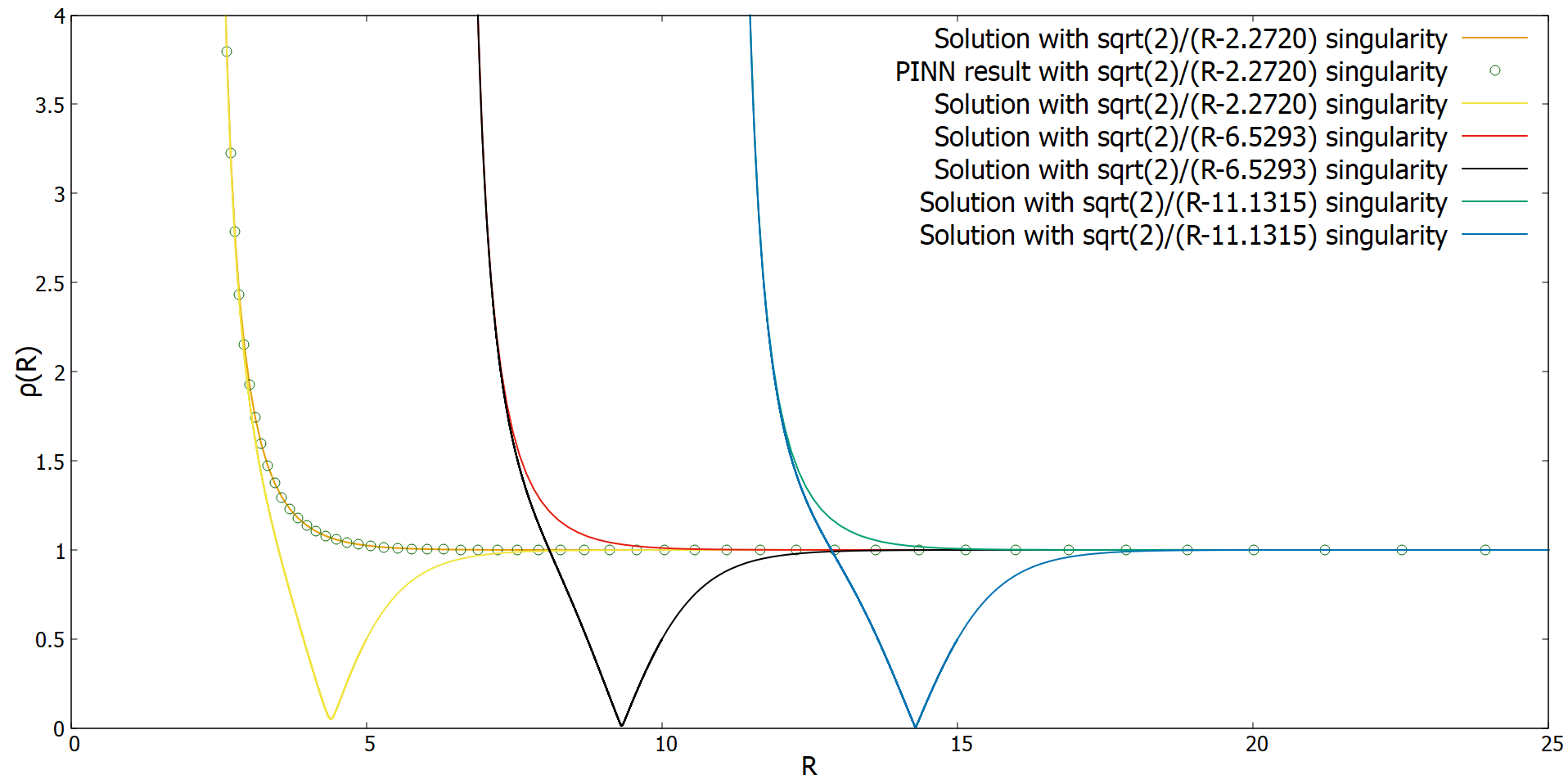}
        \caption{In 3D} \label{fig:nonunique_b}
    \end{subfigure}
    \vspace{5pt}
    \caption{Sample non-unique solutions for $\lvert B\rvert=1$. The solutions with and without the dip, represent two different families of solutions.}
    \label{fig:2D_and_3D_nonunique}
\end{figure}

\subsection{Using Newton iteration and Runge-Kutta method}
The ODE is solved for each value of $B$ using Newton iteration \cite{GALANTAI200025} with Chebyshev polynomial collocation as follows. Writing the equation in the generic form 
%\begin{equation} 
\begin{align}\label{e:StationaryODE=G=0}
ODE & = G  \left(f  \left(U \right),  f'  \left(U \right), f''  \left(U \right)\right) = 0 \nonumber \\ 
    & \text{ where}, f'  \left(U \right) = \frac{d f  \left(U \right)}{d U}, f''  \left(U \right) = \frac{d^{2} f  \left(U \right)}{d U^{2}} 
\end{align}
%\end{equation}
we expand $G$ in~\eqref{e:StationaryODE=G=0} about $f_\text{old}(U)$ at first order  \cite{na1979computational}
{\scriptsize\begin{equation} \label{e:StationaryODETaylorexpandG}
G_{old} + (f_\text{new} - f_\text{old}) \left(\left.\frac{d G}{d f}\right\vert_{f=f_\text{old}} \right) + (f_\text{new}' - f_\text{old}') \left(\left.\frac{d G}{d f'}\right\vert_{f'=f_\text{old}'} \right) + (f_\text{new}'' - f_\text{old}'') \left(\left.\frac{d G}{d f''}\right\vert_{f''=f_\text{old}''} \right) = 0
\addtocounter{equation}{1}\tag*{\normalsize(\theequation)}
\end{equation}}
and solve this to find an improved solution $f_\text{new}$. Starting from an initial guess for the function and $R_{0}$, we use this Newton iteration until $G\approx0$ for some $R_{0}\ge 0$ within the numerical approximation. Here we use Chebyshev collocation differential matrices \cite{SALEHTAHER20134634} to discretize the derivatives. These are obtained by first writing the function in the basis of Chebyshev polynomials of the first kind and degree N on the collocation points $U_k = \cos\left(\frac{\pi k}{N} \right)$ where $k=0,1,2,\ldots,N$. Then, using the properties of derivatives of these polynomials, the differential matrices are obtained. Examples of the numerical solutions obtained in 2D and 3D are shown in Figures \ref{fig:2D_numerical} and \ref{fig:3D_numerical}, respectively. The solutions obtained with this numerical approach give a lower bound on $R_0$ for a given value of B. Here, we compute the lower bound on $R_0$ with four significant figures. Note that this is only a lower bound, which means that for the same value of parameter $B$, we have infinitely many solutions that behave in a similar way asymptotically, but with singularity occurring at any $R_0 >$ this lower bound. 

There exists another family of singular solutions with singularity of the form $\frac{\sqrt{2}}{R-R_0}$ at $R_0 > 0$ as shown\customfootnote{The solutions with a bigger dip in the figures} in Figures \ref{fig:2D_numerical} and \ref{fig:3D_numerical}. We suspected the existence of this other family of solutions because of the way the transseries resummation results with negative values of the free parameter behaved. This is discussed in detail in the Laplace-Borel analysis section. Here, we obtain the corresponding solutions numerically by first solving the BVP (Boundary Value Problem) with $\rho = \frac{1}{2}$ (which is chosen arbitrarily) at the left boundary (at some $R=X$ chosen arbitrarily) and $\rho = 1$ at the right boundary (at $R\to\infty$). We solve this BVP by mapping $X \le R < \infty$ to $-1\le U \le 1$ so that we can use the Newton iteration with Chebyshev collocation. Then we use this solution to solve IVP (Initial Value Problem) starting from $R=X$ (or some $R>X$) using (RK4) the Runge-Kutta method of fourth order \cite{Hairer2015}. We try this for various values of $X$ such as $X=5, 10, 15$. It turns out that a similar procedure can be used for the logarithmic singular solution \big(for this we use $\rho  \left(X \right) = \frac{1}{1.01}$ and $X = 0.9604, 1.0000$\big) in 2D (see \eqref{e:ln(R)}) as shown in Figure \ref{fig:2D_numerical_ln}. The reason this (BVP followed by IVP) works is because the differentiation matrices of Chebysev collocation help get the derivatives of the function everywhere including boundaries of BVP with good accuracy.

Note that for any given value of $\lvert B\rvert$, the numerical calculations clearly show the existence of multiple singular solutions that all asymptotically go to $1$, as can be seen for $\lvert B\rvert = 1$ in Figures \ref{fig:nonunique_a} and \ref{fig:nonunique_b} for 2D and 3D, respectively. We try to understand this phenomenon using asymptotic transseries and resurgence in later sections.

\subsection{Using Physics Informed Neural Networks (PINN)}
This problem can also be solved using Neural Networks. Solving a nonlinear second order ODE with Dirichlet boundary conditions, using a Neural Network, is essentially an optimization problem that tries to minimize the loss function (that has to be $0$) defined as 
\beq
\sum_{i=0}^{i=N} (\mathrm{ODE} \left(f  \left(U_i \right),  f'  \left(U_i \right), f''  \left(U_i \right), B, R_0\right))^2
\eeq
where, $f'  \left(U_i \right) = \frac{d f  \left(U \right)}{d U}\Bigr|_{U_i}, f''  \left(U_i \right) = \frac{d^{2} f  \left(U \right)}{d U^{2}}\Bigr|_{U_i}$, and $\mathrm{U}_{i}$ is an $i^{th}$ grid point on coordinate U grid. Here, $\mathrm{ODE}$ is the equation \eqref{e:StationaryODE} after the substitution $\rho(U) = \frac{1}{f(U)}$ and $R \in [R_{0},\infty) \to U \in [-1,1]$ similar to Newton method with collocation as discussed earlier.

What makes this different is that instead of discretizing the derivatives using finite difference or polynomial collocation, it uses what is known as automatic differentiation. This works by applying weights and biases followed by the activation function (ReLU, Sigmoid, tanh, etc.) to the grid points repeatedly layer by layer in the Neural Network \cite{DUFERA2021100058}, and then differentiating with respect to the grid points using the chain rule. In some sense this is similar to polynomial collocation, except that the grid points, weights, and biases can be completely arbitrary unlike the basis functions and grid points used for the collocation. To demonstrate this approach, we solve this problem for $f(U)=\frac{1}{\rho(U)}$ in coordinate U with $B = 1.0$ and $R_0 = 1.6643$ (in 2D) or $R_0 = 2.2720$ (in 3D), and compare it with the result of a standard numerical technique such as Newton iteration with Chebyshev polynomial collocation as shown in figures \ref{fig:nonunique_a} and \ref{fig:nonunique_b}. For that, we set up a neural network for all grid points\customfootnote{Although they can be arbitrary, we choose the same Chebyshev collocation points as earlier for comparison with Newton method result.} with three intermediate layers containing 15, 20 and 15 neurons, respectively, and input and output layers containing one neuron each. The output of the neural network $N(U, P)$ (where $U$ are grid points and $P$s are random weights and biases) is scaled as $(1-U^2) N(U, P) + 0.5 (1+U)$ to ensure that the Dirichlet boundary conditions on $f(U)$ are satisfied. We solve for this scaled output at all grid points including boundaries, using the $tanh$ activation function and Adam optimizer \cite{DUFERA2021100058}, on the CUDA platform on NVIDIA GeForce RTX 2060 using Python library called PyTorch which includes automatic differentiation engine called autograd. These calculations serve as an independent check for the Newton iteration with Chebyshev collocation. 

\iffalse
\begin{figure}[H]
    \centering
    \includegraphics[width=\linewidth]{images_with_better_legends/NN_Newton_comparison.png}
    \caption{Comparison of solutions for $|B|=2.5$ with $R_0=2.2258$ in 2D} \label{fig:NN_Newton_comparison}
\end{figure}
\fi

Here we have used only the grid points as input variables. Essentially, the neural network is being trained for these grid points. In principle, we can input test data, i.e. other grid points than the ones already used, and get predictions for the function at those test data points. We could also include the parameters $B$ and $R_0$ (in addition to the grid points) in the input variable vector to the Neural Network and from the trained Neural Network for certain grid points, $B$, and $R_0$, we can get the predictions for the solution/function for different values of these parameters. However, we leave this last part for future work. It is important to point out that here we have used only $100$ grid points (sample size) for training, and yet the result agrees pretty well with the previous numerical solutions.

\section{Transseries solution}

Although numerical integration is useful, it is always important to use an analytic method such as series expansion to gain a deeper understanding of the solutions. In this section we analyze what type of series expansion is required to solve the differential equation. Initially we try a series solution \big($\lim_{R \to \infty} \rho  \left(R \right) = 1$\big) for this ordinary differential equation~\eqref{e:StationaryODE}. In the $\mathbf{d}=2$ case, we get a series\customfootnote{For clarity, in this section we show a few terms of the series expansions but the actual calculations where done with a computer algebra program (Maple) with up to order $\mathcal{O}(R^{-201})$.}:
\begin{equation} \label{e:2Drho}
%\begin{align}
\rho \left(R \right) = 1-\frac{B^{2}}{2 R^{2}}-\frac{\left(\frac{9}{8} B^{4}+B^{2}\right)}{R^{4}}-\frac{\left(\frac{65}{16} B^{6}+12 B^{4}+8 B^{2}\right)}{R^{6}}- \ldots\\
%\end{align}
\end{equation}

Notice first that the series is unique, there is no constant of integration. Moreover, the series has zero radius of convergence. Indeed, as can be seen 
\begin{figure}[H]
    \centering
    \begin{subfigure}{1\textwidth}
        \centering
        \includegraphics[width=\linewidth]{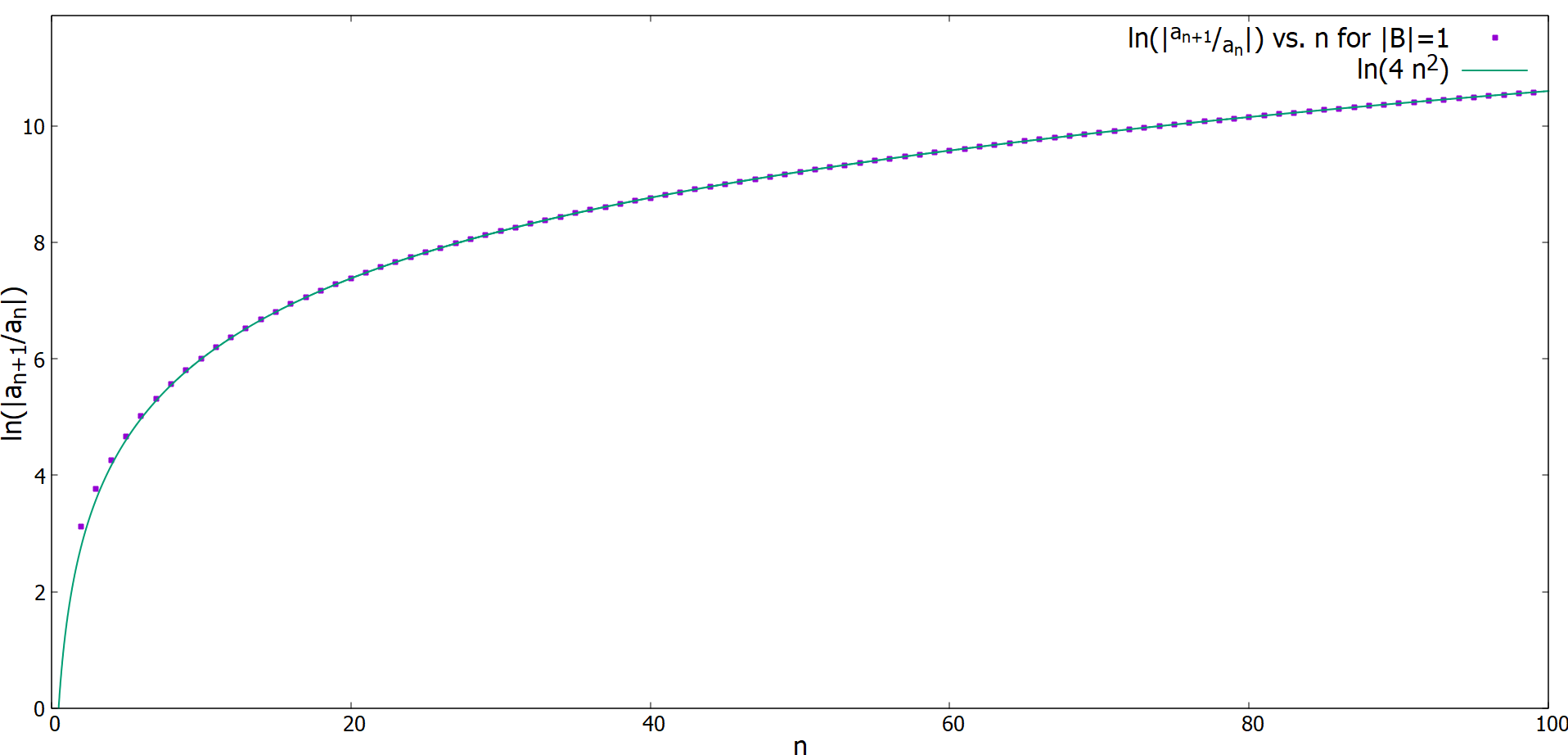}
        \caption{$\lvert B\rvert= 1$} \label{fig:B1_coefficient_growth}
    \end{subfigure}
    \\[10pt]
    \begin{subfigure}{1\textwidth}
        \centering
        \includegraphics[width=\linewidth]{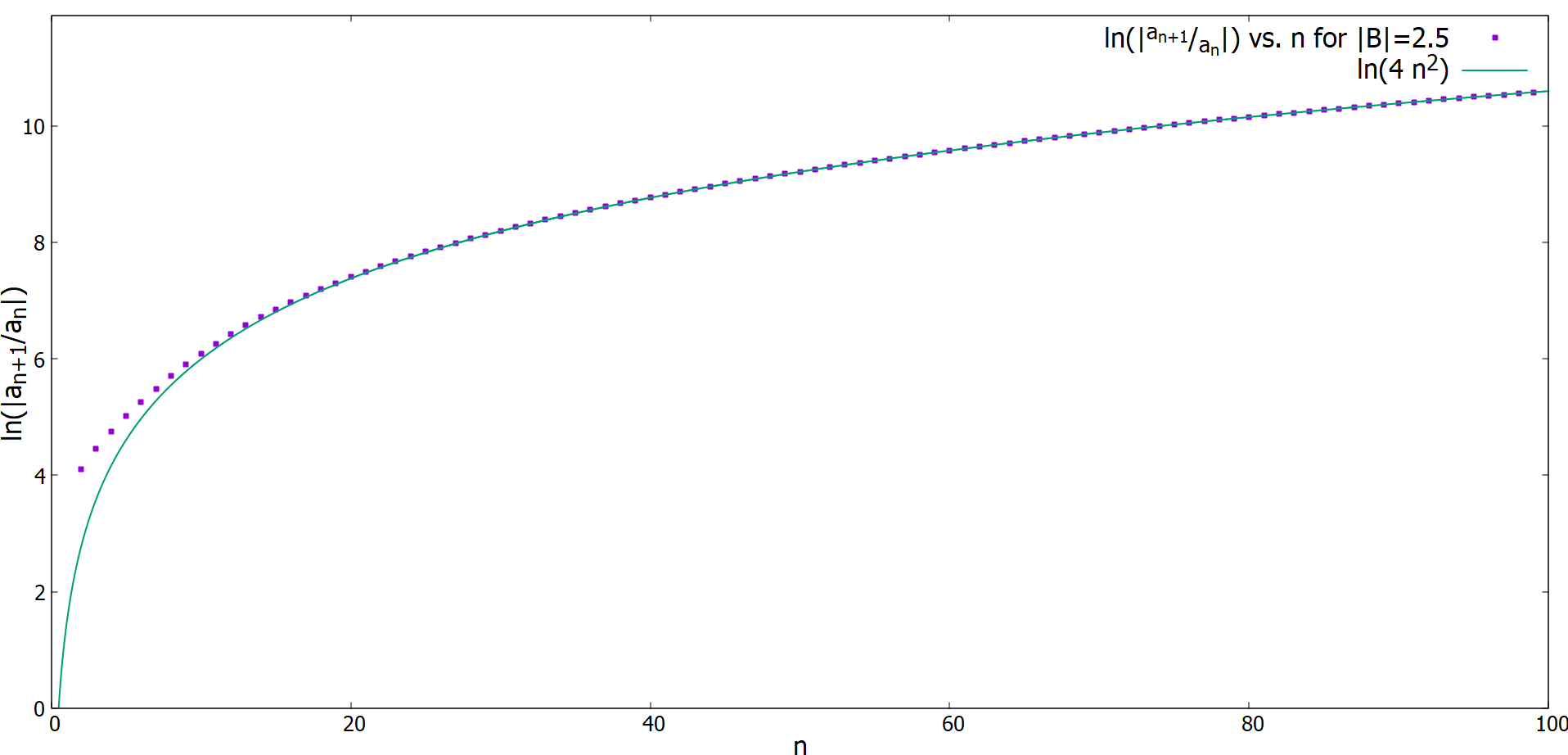}
        \caption{$\lvert B\rvert = 2.5$} \label{fig:B2.5_coefficient_growth}
    \end{subfigure}
    \vspace{5pt}
    \caption{Sample plots of $\ln\left(\left|\frac{a_{n+1}}{a_n}\right|\right)$ vs. $n$ for Stationary solutions in 2D}
    \label{fig:coefficient_growth_Stationary}
\end{figure}
\noindent in Figures \ref{fig:B1_coefficient_growth} and \ref{fig:B2.5_coefficient_growth}, at large-order $\left|\frac{a_{n+1}}{a_n}\right| \approx \frac{(2(n+1))!}{(2n)!} \to 4 n^2$. In such a case, as described in the next section, we can use Laplace-Borel resummation \cite{Costin_2019} to obtain a well-defined function that solves the differential equation. Unfortunately, the Laplace-Borel transform has an imaginary part in contradiction with the physical requirement that $\rho(R)$ is real. The absence of an arbitrary constant of integration and a nonzero imaginary part in the resummation, indicate the existence of asymptotic non-perturbative (non-power law) corrections to this series expansion. To check this, we first substitute $\rho  \left(R \right) = 1 + f(R)$ in~\eqref{e:StationaryODE} and linearize in $f(R)$, (see \cite{Costin2015} and \cite{10.1155/S1073792895000286}) which gives $\rho  \left(R \right) \approx 1 + A {R}^{-\frac{(\mathbf{d}-1)}{2}} \exp(-\sqrt{2}  R) + \ldots$ where $A$ is an arbitrary constant of integration for the linearized ODE. Note that $\exp(+\sqrt{2}  R)$ non-perturbative corrections are discarded since we are looking for solutions (on $R\geq0$) that approach $1$ as $R\to\infty$.

Making a change of variable $u = \frac{1}{\sqrt{2}  R}$ in~\eqref{e:StationaryODE}, for $\mathbf{d}=2$ we get 
\begin{equation} \label{e:2DStationaryODEcoordinate_u}
%\begin{align}
\frac{d^{2} \rho }{d u^{2}}+\frac{1}{u} \frac{d \rho  }{d u}+\frac{\rho -\rho  ^{3}}{2 u^4}-\frac{B^{2}}{\rho  ^{3} u^{2}} = 0
%\end{align}
\end{equation}

We now substitute a transseries ansatz \eqref{e:transseries_ansatz} in \eqref{e:2DStationaryODEcoordinate_u}
\beq \label{e:transseries_ansatz}
\rho  \left(u \right) =  \sum_{j=0}^\infty f_{j} \left(u \right) {\mathrm e}^{-\frac{j}{u}} {u}^{\frac{(\mathbf{d}-1) j}{2}} c^{j}
\eeq
where we introduced an arbitrary constant $c = \sigma_{\Re} + i  \sigma_{\Im}$. The differential equation \eqref{e:2DStationaryODEcoordinate_u} and transseries \eqref{e:transseries_ansatz} lead to a series of differential equations for the $f_j(u)$. In 2D those equations have the formal series solutions:
%\begin{equation} 
\begin{flalign} \label{e:2Df[i]s}
f_{0} \left(u \right) & = 1-B^{2} u^{2}+\left(-\frac{9}{2} B^{4}-4 B^{2}\right) u^{4}+\left(-\frac{65}{2} B^{6}-96 B^{4}-64 B^{2}\right) u^{6}+ \ldots, \nonumber\\
f_{1} \left(u \right) & = 1+\left(-3 B^{2}-\frac{1}{8}\right) u +\left(\frac{9}{2} B^{4}+\frac{15}{8} B^{2}+\frac{9}{128}\right) u^{2}+\ldots, \nonumber\\
f_{2} \left(u \right) & = \frac{1}{2}+\left(-3 B^{2}-\frac{11}{24}\right) u +\left(9 B^{4}+\frac{35}{4} B^{2}+\frac{53}{64}\right) u^{2}+\ldots, \nonumber\\
f_{3} \left(u \right) & = \frac{1}{4}+\left(-\frac{9 B^{2}}{4}-\frac{13}{32}\right) u +\left(\frac{81}{8} B^{4}+\frac{233}{32} B^{2}+\frac{461}{512}\right) u^{2}+\ldots, \nonumber\\
f_{4} \left(u \right) & = \frac{1}{8}+\left(-\frac{3 B^{2}}{2}-\frac{7}{24}\right) u +\left(9 B^{4}+\frac{53}{8} B^{2}+\frac{145}{192}\right) u^{2}+\ldots \nonumber\\
\end{flalign}
%\end{equation}

We only show the first few terms for the $f_j(u)$ up to $j=4$. However, we use up to $j=10$ with each $f_{j}(u)$ series up to $\mathcal{O}\left(u^{201}\right)$ for the actual calculations. Note that $f_{0}(u)$ is exactly the same as \eqref{e:2Drho} with $u=\frac{1}{\sqrt{2}R}$. Given that all these series have zero radius of convergence, to obtain an actual solution requires a series of further steps. First we use Borel resummation to obtain functions $F_j(u)$ that satisfy the same differential equations as $f_j(u)$ thus obtaining a function $\rho(u)$ in \eqref{e:transseries_ansatz} that satisfies the original  equation \eqref{e:2DStationaryODEcoordinate_u}. As already mentioned such function is generically not real. To obtain a real solution we have to choose appropriately the imaginary part $\sigma_\Im$ of the constant $c$ in the transseries. This is sufficient due to the phenomenon of resurgence. After all that is done we obtain a concrete solution that we can evaluate and compare with the numerical results. The real part $\sigma_\Re$ of the constant $c$ appears as an arbitrary constant of integration that accounts for the multiple solutions seen in \ref{fig:nonunique_a} (and \ref{fig:nonunique_b}).

% with series of $\mathrm{O} \left(u^{2  b + 1}\right)$
\subsection{Laplace-Borel Analysis}
If we solve a differential equation with a series expansion that has a finite radius of convergence, we obtain a solution on a disk and, by analytic continuation, on a larger region of the complex plane. On the other hand, factorially divergent series such as the ones for $f_j(u)$ in \eqref{e:2Df[i]s} have zero radius of convergence. In such case we can obtain a solution using Laplace-Borel resummation. First we define the Borel transforms (\cite{10.1155/S1073792895000286}, \cite{Costin_2019}) of order $\mathrm{O} \left(v^{2 b}\right)$ as\customfootnote{As we shall see, it turns out that $f_{j}(u)$ series have $\sim (2n)!$ and $\sim n!$ divergence for $j=0$ and $j>0$ respectively.}
\begin{align}
\mathcal{F}_0(v)=\mathcal{B}[\left(f_{0} \left(u \right) - \left(a_{0}\right)_{0} \right)] & = \sum_{k=1}^{b} \frac{\left(a_{0}\right)_{k}}{(2  k - 1)!}  v^{2  k-1} \nonumber \\
\mathcal{F}_{j}(v)=\mathcal{B}[\left(f_{j} \left(u \right) - \left(a_{j}\right)_{0} \right)] & = \sum_{k=1}^{2  b} \frac{\left(a_{j}\right)_{k}}{(k - 1)!}  v^{k-1} \text{ for }j>0\nonumber
\end{align}
The series $\mathcal{F}_{j}(v)$ have a finite radius of convergence and define, by analytic continuation,  solutions to the Borel transform of the original equation. Their Borel inverse (or Laplace) transform satisfy the original differential equation that $f_j(u)$ were supposed to solve. A crucial point is that to perform the Laplace transform we have to evaluate the functions $\mathcal{F}_{j}(v)$ outside the radius of convergence of the series and therefore we need a method to perform analytic continuation in $v$. Since the Borel transform here cannot be written in a closed functional form, we approximate this Borel transform using the Borel-Pad\'e approximation or the Conformal-Borel-Pad\'e approximation before performing the Laplace integration. 

\subsubsection{Borel-Pad\'e approximation}
\begin{figure}[H]
    \centering
    \begin{subfigure}{0.65\textwidth}
        \centering
        \includegraphics[width=\linewidth]{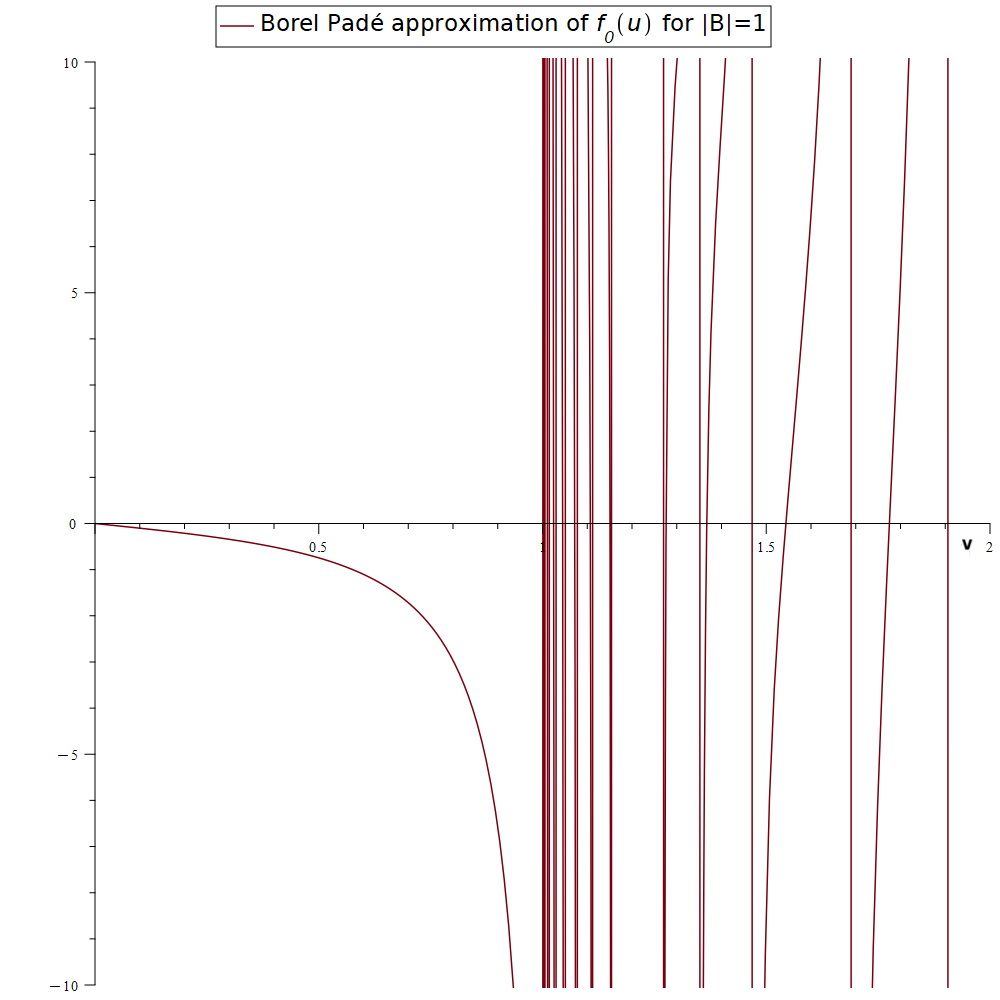}
        \caption{Borel Pad\'e of $f_{0}$ for $|B|=1$ in 2D} \label{fig:B1_f0_BP}
    \end{subfigure}
    \\[10pt]
    \begin{subfigure}{0.65\textwidth}
        \centering
        \includegraphics[width=\linewidth]{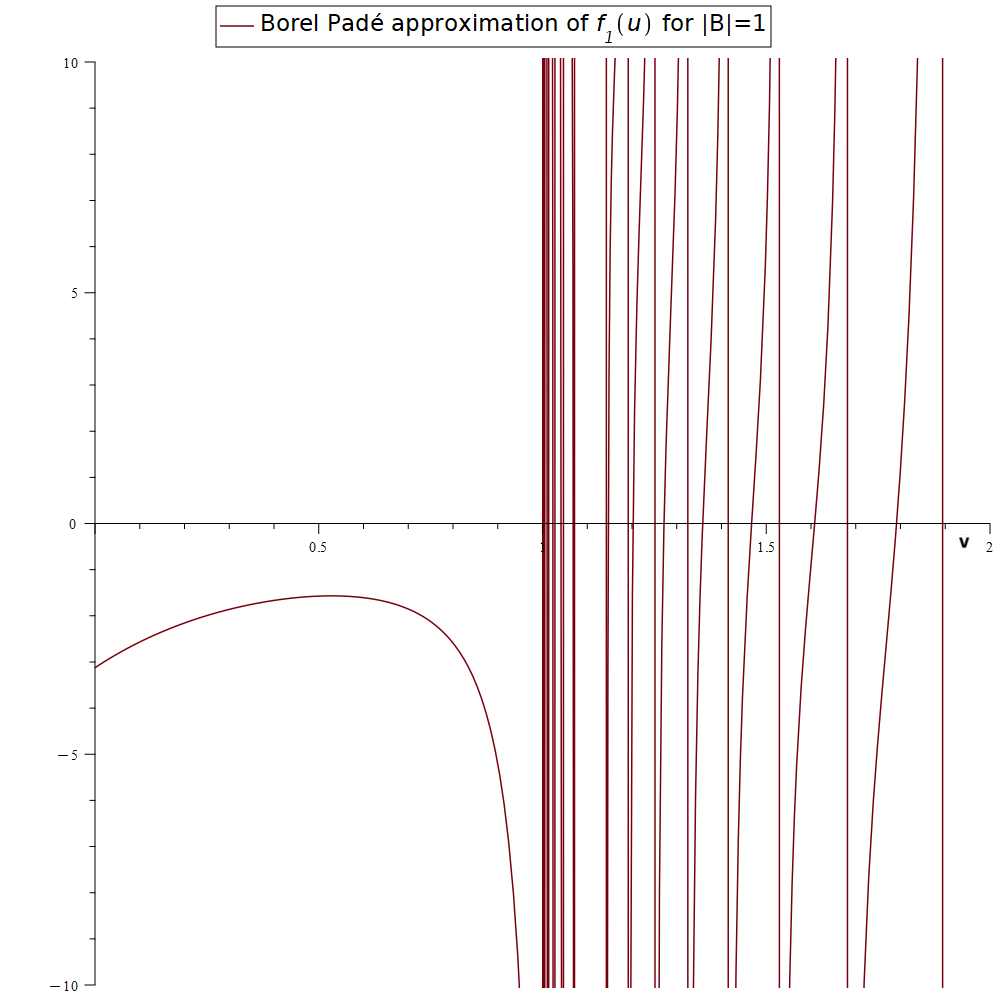}
        \caption{Borel Pad\'e of $f_{1}$ for $|B|=1$ in 2D} \label{fig:B1_f1_BP}
    \end{subfigure}
    \vspace{5pt}
    \caption{Borel Pad\'e approximation}
    \label{fig:Borel Pade approximation}
\end{figure}

\begin{figure}[H]
    \centering
    \begin{subfigure}{0.65\textwidth}
        \centering
        \includegraphics[width=\linewidth]{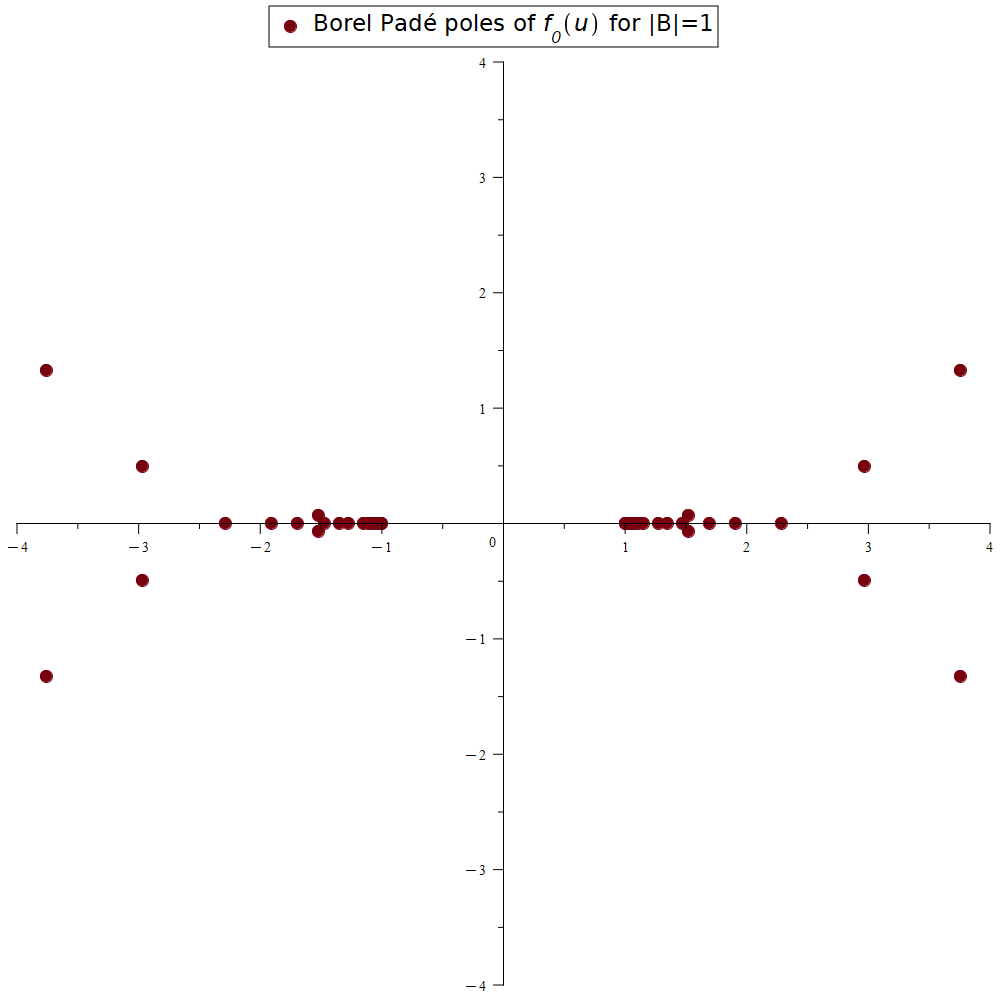}
        \caption{Approximate Borel Pad\'e poles of $f_{0}$ for $|B|=1$ in 2D} \label{fig:B1_f0_BP_poles}
    \end{subfigure}
    \\[10pt]
    \begin{subfigure}{0.65\textwidth}
        \centering
        \includegraphics[width=\linewidth]{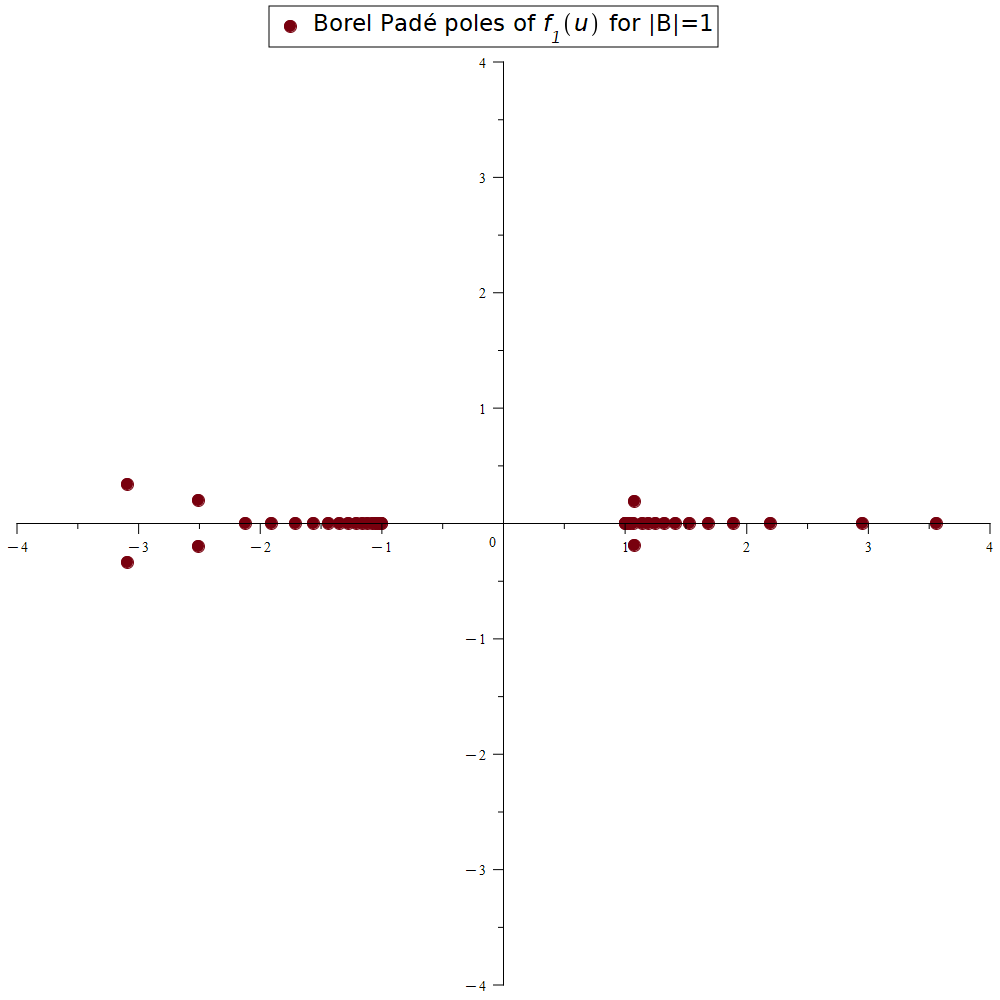}
        \caption{Approximate Borel Pad\'e poles of $f_{1}$ for $|B|=1$ in 2D} \label{fig:B1_f1_BP_poles}
    \end{subfigure}
    \vspace{5pt}
    \caption{Approximate Borel Pad\'e poles in $v$ plane}
    \label{fig:Borel Pade poles}
\end{figure}

The Pad\'e approximation $[a,b]$ approximates an analytic function $f(v)$ with a rational function $f(v)=\frac{P_a(v)}{Q_b(v)}$ where $P_a$, $Q_b$ are polynomials of order $a,b$. From the Pad\'e approximation of order $[b-1,b]$ of the Borel transform, it can be seen that the leading Borel singularity at $+1$ in the Borel plane falls on the Laplace integration path (the real line) as shown in Figure~\ref{fig:Borel Pade approximation} and~\ref{fig:Borel Pade poles} for $\lvert B \rvert = 1$. Therefore, lateral Laplace-Borel resummation is performed by doing the Laplace integral in the Borel plane just above or just below the cut as follows:

\begin{equation}
\text{Resummed } f_{j} \left(u \right) \implies F_{j}(u) = \left(a_{j}\right)_{0} + \int_{0}^{\infty +\mathrm{i}  {0}^{\pm}}{\mathrm e}^{-\frac{v}{u}} \mathcal{P}[\mathcal{F}_{j} \left(v \right)] d v \nonumber
\end{equation}
where, $j = 0 \ldots 10$. Here, $\mathcal{P}[\mathcal{F}]$ is a Pad\'e approximation of order [b-1,b] of the Borel transform $\mathcal{F}$ of order $\mathrm{O} \left(v^{2 b}\right)$.

Assuming that the leading singularity of this function on the Laplace integration path is of the form $\frac{\alpha}{(1-v)^\lambda}$, we find

\begin{equation} \label{e:LeadingSingularity}
%\begin{align}
\lambda = \lim_{v \rightarrow 1^{-}} \frac{1}{(\mathcal{P}[\mathcal{F}_j \left(v \right)])} \frac{d (\mathcal{P}[\mathcal{F}_j  \left(v \right)])}{d v} (1-v) \approx \frac{1}{2} \text{ ,} \hspace{10pt} \forall \hspace{5pt} j
%\end{align}
\end{equation}
suggesting that the leading singularity is a square root branch point.

As can be seen in figures \ref{fig:Borel Pade approximation} and \ref{fig:Borel Pade poles}, Borel-Pad\'e approximation has spurious poles beyond $\pm1$ on the real line in the Borel plane (similar to \cite{PhysRevD.92.125011}). Although this appears problematic, we can still perform an approximate lateral Laplace-Borel transform just below (or just above) the real line. Later we take $\sigma_{\Im}>0$, corresponding a Laplace integral below the real axis. The resulting function is complex. If the equation were linear we could obtain a real solution by averaging the integrals above and below the real axis. Since it is not linear we have to use resurgence which effectively turns out to be a median summation \cite{Aniceto2015}. Before that, however, we improve on the Pad\'e approximation by using a conformal map.

\iffalse
\begin{figure}[H]
    \centering
    \includegraphics[width=150mm,scale=0.5]{images_with_better_legends/deformed_Laplace_contour_integral_avoiding_singularity.png}
    \caption{Deformed contour avoiding singularity for $\mathcal{LBP}$ and $\mathcal{LCBP}$}
    \label{fig:Contour avoiding singularity for LBP and LCBP}
\end{figure}
\fi

\subsubsection{Conformal-Borel-Pad\'e approximation}
An improvement on the approximation with rational functions is Conformal-Borel-Pad\'e approximation. It can be obtained as follows   \cite{Costin_2019}:
\begin{itemize}
\item[1)] Substitute $v = \frac{2  z}{1+z^{2}}$ into the Borel transform.
\item[2)] Expand it inside the unit disc about z=0 up to $\mathrm{O} \left(z^{2  b}\right)$ and convert it into a Pad\'e approximation of order $[b-1,b]$ in the variable $z$.
\item[3)] Substitute back $z = \frac{v}{1+\sqrt{1-v^{2}}}$
\end{itemize}

\begin{figure}[H]
    \centering
    \begin{subfigure}{0.65\textwidth}
        \centering
        \includegraphics[width=\linewidth]{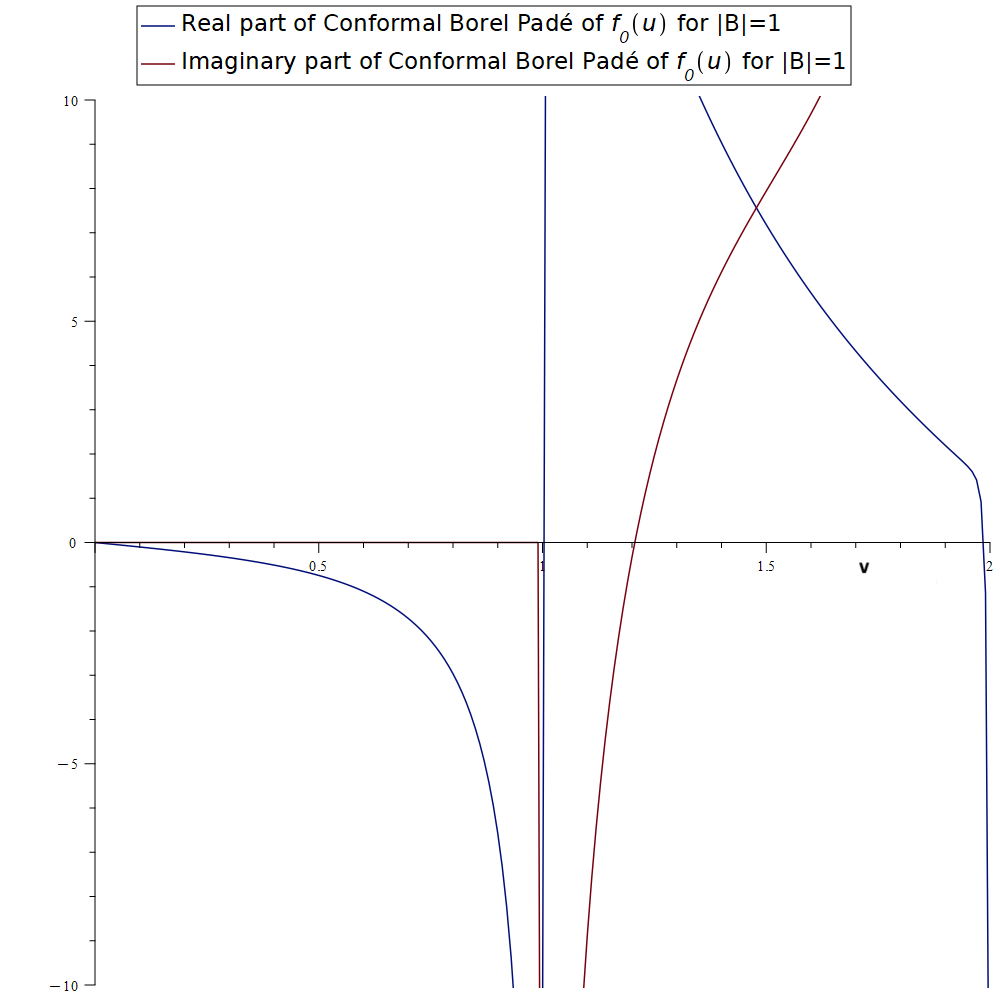}
        \caption{Conformal Borel Pad\'e of $f_{0}$ for $|B|=1$ in 2D} \label{fig:B1_f0_CBP}
    \end{subfigure}
    \\[10pt]
    \begin{subfigure}{0.65\textwidth}
        \centering
        \includegraphics[width=\linewidth]{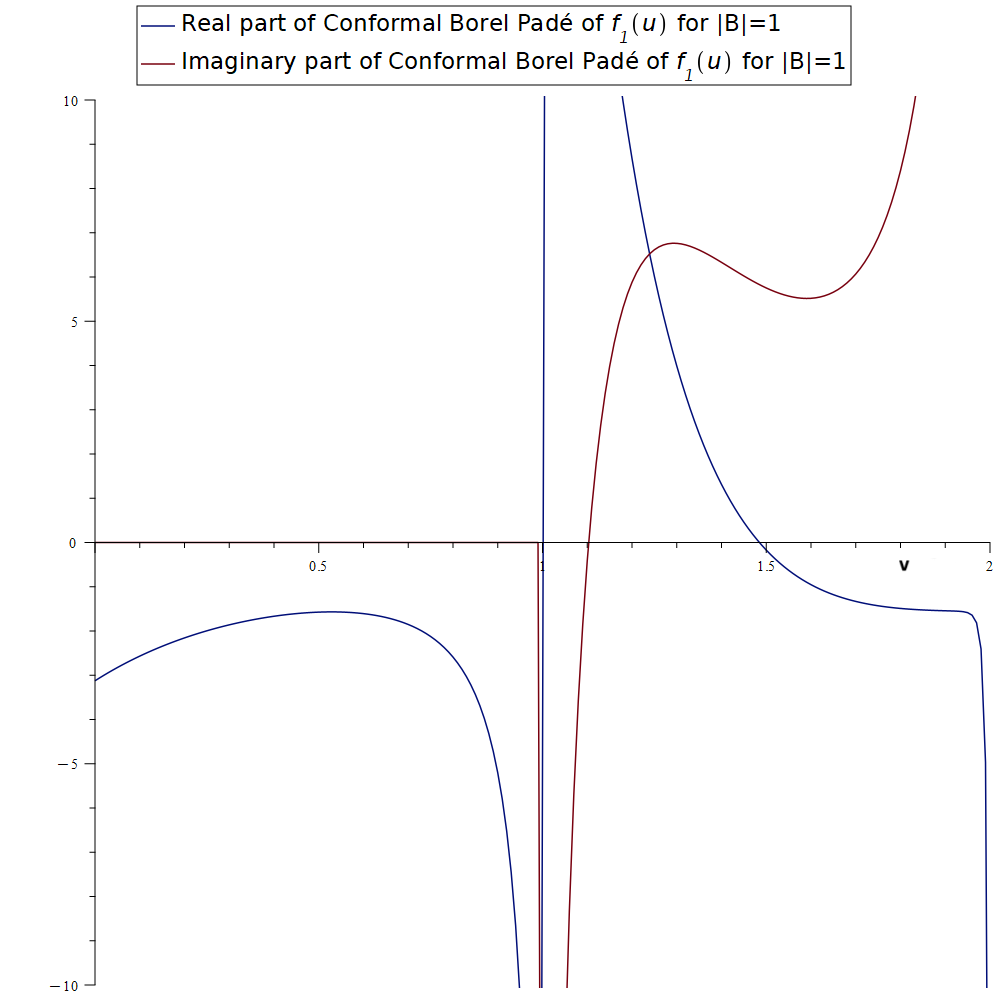}
        \caption{Conformal Borel Pad\'e of $f_{1}$ for $|B|=1$ in 2D} \label{fig:B1_f1_CBP}
    \end{subfigure}
    \vspace{5pt}
    \caption{Real and Imaginary parts of Conformal Borel Pad\'e approximation}
    \label{fig:Conformal Borel Pade approximation}
\end{figure}

After conformal mapping of the functions shown in Figure \ref{fig:Borel Pade approximation}, we get the Conformal Borel Pad\'e approximation as shown in Figure \ref{fig:Conformal Borel Pade approximation}. The Laplace transform gives:
\begin{equation}
\text{Resummed } f_{j} \left(u \right) \implies F_{j}(u) = \left(a_{j}\right)_{0} + \int_{0}^{\infty +\mathrm{i}  {0}^{\pm}}{\mathrm e}^{-\frac{v}{u}} \mathcal{CP}[\mathcal{F}_{j} \left(v \right)] d v \nonumber
\end{equation}
where, in practice, we only keep  $j = 0 \ldots 10$. Here,  $\mathcal{CP}[\mathcal{F}]$ is a Pad\'e approximation of order [b-1,b] of the conformally mapped Borel transform $\mathcal{F}$ of order $\mathrm{O} \left(z^{2 b}\right)$ wherein we substitute $z$ in terms of $v$ after this procedure.

Since the Borel singularity (in Conformal Borel-Pad\'e approximation) in the coordinate $z$ is at $1$, we Pad\'e approximate the functions as (using that $f_0(u)$ is even in $u$):
\begin{align} \label{e:CBP expansion}
\frac{\mathcal{P}[(1-z^2) \mathcal{C}[\mathcal{F}_{0}(v)]}{1-z^2} &= \frac{A_{0}}{1-z} + \tilde{F}_0(z) \nonumber \\
\frac{\mathcal{P}[(1-z) \mathcal{C}[\mathcal{F}_{j}(v)]]}{1-z} &= \frac{A_{j}}{1-z} + \tilde{F}_{j}(z) \text{ for $j>0$}\nonumber \\
\end{align}
where $\cal{P}$ are rational polynomials in $z$ of order $[b,b+1]$ and $[b+1,b+2]$ respectively. 

\iffalse
\begin{figure}[H]
    \centering
    \includegraphics[width=.50\textwidth]{images_with_better_legends/B1_LBP_LCBP_epsilon_1e-2_singularity_at10.png}\hfill
    \includegraphics[width=.50\textwidth]{images_with_better_legends/B1_LBP_LCBP_epsilon_1e-2_singularity_at10_zoomed.png}
    \vspace{5pt}
    \caption{$\frac{1}{\rho  \left(R \right)}$ obtained using Newton iteration, and lateral Laplace-Borel transform with and without conformal mapping, with 4, 7, and 10 exponential terms respectively, for $|B|$=1 with singularity at R=10 in 2D}\label{fig:2D_stationary_resummation}
\end{figure}
\fi

\iffalse
\begin{figure}[H]
\centering
    \includegraphics[width=\linewidth]{images_with_better_legends/B1_InvRho_how_good_resummation_is_Other_family_of_sol.png}
    \vspace{5pt}
    \caption{A sample solution for $\frac{1}{\rho  \left(R \right)}$ in 2D with $|B| = 1$, obtained using BVP (Boundary Value Problem)+IVP (Initial Value Problem), and matched with the resummation result}\label{fig:2D_stationary_resummation2}
\end{figure}

\begin{figure}[H]
\centering
    \includegraphics[width=\linewidth]{images_with_better_legends/B1_Rho_how_good_resummation_is_Im_Other_family_of_sol.png}
    \vspace{5pt}
    \caption{Imaginary part of transseries without resummation and resummed transseries, for the other family of solution \ref{fig:2D_stationary_resummation2} with singularity at $R_{0} \sim 1.6643$} \label{fig:Im richness of resummation for the other family}
\end{figure}
\fi
The singularity at $z=1$ is then easily extracted determining the numerical constants $A_0$, $A_j$.
The Laplace transform of the non-singular parts $\tilde{F}_0$, $\tilde{F}_j$ is done numerically and the transform of $\frac{1}{1-z}$ can be done by a combination of numerical and analytical integration.
\iffalse
\begin{figure}[H] 
\centering
    \includegraphics[width=\linewidth]{images_with_better_legends/B1_InvRho_how_good_resummation_is.png}
    \vspace{5pt}
    \caption{Inverse of numerical solution, transseries without resummation, and resummed transseries} \label{fig:richness of resummation1}
\end{figure}

\begin{figure}[H] 
\centering
    \includegraphics[width=\linewidth]{images_with_better_legends/B1_Rho_how_good_resummation_is_Im.png}
    \vspace{5pt}
    \caption{Imaginary part of transseries without resummation and resummed transseries for the solution \ref{fig:richness of resummation1} with singularity at $R_{0} \sim 10.5303$} \label{fig:Im richness of resummation1}
\end{figure}
\begin{figure}[H]
\centering
    \includegraphics[width=\linewidth]{images_with_better_legends/B1_InvRho_how_good_resummation_is_Other_family_of_sol.png}
    \vspace{5pt}
    \caption{A sample solution for $\frac{1}{\rho  \left(R \right)}$ in 2D with $|B| = 1$, obtained using BVP (Boundary Value Problem)+IVP (Initial Value Problem), and matched with the resummation result}\label{fig:2D_stationary_resummation2}
\end{figure}

\begin{figure}[H]
\centering
    \includegraphics[width=\linewidth]{images_with_better_legends/B1_Rho_how_good_resummation_is_Im_Other_family_of_sol.png}
    \vspace{5pt}
    \caption{Imaginary part of transseries without resummation and resummed transseries, for the other family of solution \ref{fig:2D_stationary_resummation2} with singularity at $R_{0} \sim 1.6643$} \label{fig:Im richness of resummation for the other family}
\end{figure}
\fi
The resummation results agree well with the numerical results\customfootnote{The comparison requires the value of $\sigma_{\Im}$ which is obtained using resurgence as discussed in the next subsection.}. Figure~\ref{fig:richness of trans-asymptotic1} contains a sample solution for $\rho  \left(R \right)$ with $|B| = 1$ and singularity at $R_0 = 10.5303$, to show how well the resummation of the asymptotic resurgent transseries agrees with the numerical solution as opposed to the same transseries without resummation. Figure~\ref{fig:Im richness of trans-asymptotic1} shows that the imaginary part of the resummed transseries is approximately $0$ in the range of agreement, while the imaginary part of the transseries without the resummation diverges very quickly in this range. We can also see that negative values of $\sigma_{\Re}$ in the resummed transseries provide the other family of solutions which were obtained numerically. The agreement is shown in Figure \ref{fig:richness of trans-asymptotic1} (with singularity at $R_0 = 1.6643$) and its imaginary part in~\ref{fig:Im richness of trans-asymptotic1}. Similarly, we matched all the numerical solutions discussed in the previous sections with the resummation results. The approximate values of $\sigma_{\Re}$ in 2D for different types of singularity at different locations are shown in \autoref{table:sigma_Re_values}. 
\begin{table}[H]
\centering
\begin{tabular}{||p{2cm}|p{2cm}|p{2cm}|p{4cm}||}
\hline
\large $\boldsymbol{\sigma_{\Re}>0}$ & \large $\boldsymbol{\sigma_{\Re}<0}$ & \large $\boldsymbol{R_{0}}$ & \textbf{\large Singularity type} \\ [0.5ex] 
\hline
1.5 &  &  & $\frac{1}{R},\ \ \ \ \ \ \ R\rightarrow 0$ \\ 
\hline
& -16.6 &  & $-\ln{R},\ \ R\rightarrow 0$ \\
\hline
& -10.1 &  & $-\ln{R},\ \ R\rightarrow 0$ \\
\hline
5.60E+01 & -1.54E+03 & 1.2991 & $\frac{\sqrt{2}}{R-R_{0}},\ \ \  R\rightarrow R_0$ \\
\hline
9.10E+01 & -3.15E+03 & 1.6643 & $\frac{\sqrt{2}}{R-R_{0}},\ \ \  R\rightarrow R_0$ \\ [1ex] 
\hline
3.55E+04 & -4.28E+06 & 5.9077 & $\frac{\sqrt{2}}{R-R_{0}},\ \ \  R\rightarrow R_0$ \\ [1ex] 
\hline
2.80E+07 & -5.80E+09 & 10.5303 & $\frac{\sqrt{2}}{R-R_{0}},\ \ \  R\rightarrow R_0$ \\ [1ex] 
\hline
\end{tabular}
\vspace{1pt}
\caption{Approximate values of $\sigma_{\Re}$ in 2D for different types of singularities.}\label{table:sigma_Re_values}
\end{table}

\iffalse
\begin{figure}[H] 
\centering
    \includegraphics[width=\linewidth]{images_with_better_legends/B1_InvRho_how_good_resummation_is.png}
    \vspace{5pt}
    \caption{Inverse of numerical solution, transseries without resummation, and resummed transseries} \label{fig:richness of resummation1}
\end{figure}

\begin{figure}[H] 
\centering
    \includegraphics[width=\linewidth]{images_with_better_legends/B1_Rho_how_good_resummation_is_Im.png}
    \vspace{5pt}
    \caption{Imaginary part of transseries without resummation and resummed transseries for the solution \ref{fig:richness of resummation1} with singularity at $R_{0} \sim 10.5303$} \label{fig:Im richness of resummation1}
\end{figure}
\fi

\subsection{Resurgence}

After doing the Borel resummation, we obtained a resummed transseries with real and imaginary parts given by:
%\begin{subequations} 
\begin{align} %\label{e:Re and Im parts of resummed transseries}
\label{e:Re part of resummed transseries} 
\Re\left(\rho  \left(u \right) \right) &= \Re\left(F_{0}\left(u \right)\right)+{\mathrm e}^{-\frac{1}{u}} \sqrt{u}\, \left(-\sigma_{\Im} \Im\left(F_{1}\left(u \right)\right)+\sigma_{\Re} \Re\left(F_{1}\left(u \right)\right)\right)+\nonumber\\
& {\mathrm e}^{-\frac{2}{u}} u \left(-\sigma_{\Im}^{2} \Re\left(F_{2}\left(u \right)\right)-2 \sigma_{\Im} \sigma_{\Re} \Im\left(F_{2}\left(u \right)\right)+\sigma_{\Re}^{2} \Re\left(F_{2}\left(u \right)\right)\right)+\ldots \\
\label{e:Im part of resummed transseries} \Im\left(\rho  \left(u \right) \right) &= \Im\left(F_{0}\left(u \right)\right)+{\mathrm e}^{-\frac{1}{u}} \sqrt{u}\, \left(\sigma_{\Im} \Re\left(F_{1}\left(u \right)\right)+\sigma_{\Re} \Im\left(F_{1}\left(u \right)\right)\right)+\nonumber\\
&{\mathrm e}^{-\frac{2}{u}} u \left(-\sigma_{\Im}^{2} \Im\left(F_{2}\left(u \right)\right)+2 \sigma_{\Im} \sigma_{\Re} \Re\left(F_{2}\left(u \right)\right)+\sigma_{\Re}^{2} \Im\left(F_{2}\left(u \right)\right)\right)+\ldots
\end{align}
%\end{subequations}
where, the reality condition on $\rho(u)$ requires $\Im\left(\rho  \left(u \right) \right)=0$.

To understand the concept of resurgence, for the moment let us consider resummation of  $f_0\left(u \right)$ in the transseries,
\begin{align} %\label{e:Re and Im parts of resummed f0 transseries}
\label{e:Re part of resummed f_0 transseries} \Re\left(\rho  \left(u \right) \right) &= \Re\left(F_0\left(u \right)\right) + \sigma_{\Re}\, {\mathrm e}^{-\frac{1}{u}} \sqrt{u} \,f_{1}\left(u \right)+\left(\sigma_{\Re}^{2} - \sigma_{\Im}^{2}\right) {\mathrm e}^{-\frac{2}{u}} u\, f_{2}\left(u \right) \nonumber\\ &+\left(-3 \sigma_{\Im}^{2} \sigma_{\Re} +\sigma_{\Re}^{3}\right) {\mathrm e}^{-\frac{3}{u}} u^{\frac{3}{2}} \,f_{3}\left(u \right)+\ldots \\
\label{e:Im part of resummed f_0 transseries} \Im\left(\rho  \left(u \right) \right) &= \Im\left(F_0\left(u \right)\right) + \sigma_{\Im}\, {\mathrm e}^{-\frac{1}{u}} \sqrt{u} \,f_{1}\left(u \right) +2 \sigma_{\Re} \sigma_{\Im}\, {\mathrm e}^{-\frac{2}{u}} u\, f_{2}\left(u \right) \nonumber\\ &+ \left(-\sigma_{\Im}^{3}+3 \sigma_{\Im} \sigma_{\Re}^{2}\right) {\mathrm e}^{-\frac{3}{u}} u^{\frac{3}{2}} \,f_{3}\left(u \right)+\ldots
\end{align}
where, $\Re\left(F_0\left(u \right)\right)$ and $\Im\left(F_0\left(u \right)\right)$ respectively are the real and the imaginary parts of Borel resummed $f_0\left(u \right)$. 

As can be seen in Figure \ref{fig:coefficient_growth_Stationary}, in 2D $(\mathbf{d}=2)$, the leading large-order behavior of terms in the perturbative series $f_{0} \left(u \right)$ is  $(2  n)!$  $u^{2  n}$ (up to a multiplying constant) where $u = \frac{1}{\sqrt2  R}$. Resurgence predicts (see \cite{f1d7d86e-5c82-323a-9807-a6be125c47d8} and \cite{Aniceto2015}) that sub-leading corrections to the leading large-order behavior of coefficients of $f_{0} \left(u \right)$ depend on the leading low-order behavior of the coefficients of $f_{j} \left(u \right)$ for j$>$0 (similar to \cite{PhysRevD.92.125011}). 
More precisely, in the 2D case, the series coefficients of $f_{0} \left(u \right)$, namely $\left(a_{0}\right)_{k}$ should behave, for large order $k$, as \begin{align}\label{e:2DStationaryODE_resurgence?}
\left(a_{0}\right)_{k} 
\sim & C_1 \left\{-\Gamma  \left(2 k -\frac{1}{2}\right) \left(1+\frac{\left(a_{1}\right)_{1}}{2 k -\frac{3}{2}}+\frac{\left(a_{1}\right)_{2}}{\left(2 k -\frac{3}{2}\right) \left(2 k -\frac{5}{2}\right)} \right.\right. \nonumber \\
& \left.\left. \ \ \ +\frac{\left(a_{1}\right)_{3}}{\left(2 k -\frac{3}{2}\right) \left(2 k -\frac{5}{2}\right) \left(2 k -\frac{7}{2}\right)} + \ldots\right) + \ldots\right\}
\end{align}
where the $(a_1)_n$ are the coefficients of $f_1(u)$ and $C_1$ is an unknown constant. To check this prediction, we take the ratio $\frac{\left(a_{0}\right)_{k+1}}{\left(a_{0}\right)_{k}}$, Taylor expand it for large $k$, and simplify it, obtaining
{\scriptsize\begin{align}\label{e:2DStationaryODE_resurgence?_taylor}
\frac{\left(a_{0}\right)_{k+1}}{4 \left(a_{0}\right)_{k}} - k^2
= & -\left(\frac{1}{16}+\frac{\left(a_{1}\right)_{1}}{2}\right)+\left(\frac{1}{4} \left(a_{1}\right)_{1}^{2}-\frac{1}{4} \left(a_{1}\right)_{1}-\frac{1}{2} \left(a_{1}\right)_{2}\right) \left(\frac{1}{k} \right) \nonumber \\ & +\left(\frac{3}{8} \left(a_{1}\right)_{1} \left(a_{1}\right)_{2}-\frac{3}{8} \left(a_{1}\right)_{3}-\frac{3}{16} \left(a_{1}\right)_{1}-\frac{3}{4} \left(a_{1}\right)_{2}+\frac{5}{16} \left(a_{1}\right)_{1}^{2}-\frac{1}{8} \left(a_{1}\right)_{1}^{3}\right) \left(\frac{1}{k} \right)^{2} \nonumber \\ & +\left(\frac{7}{8} \left(a_{1}\right)_{1} \left(a_{1}\right)_{2}+\frac{1}{4} \left(a_{1}\right)_{1} \left(a_{1}\right)_{3}-\frac{1}{4} \left(a_{1}\right)_{1}^{2} \left(a_{1}\right)_{2}-\frac{9}{8} \left(a_{1}\right)_{3} \right. \nonumber \\ & -\left.\frac{1}{4} \left(a_{1}\right)_{4}-\frac{9}{64} \left(a_{1}\right)_{1}-\frac{33}{32} \left(a_{1}\right)_{2}+\frac{21}{64} \left(a_{1}\right)_{1}^{2}-\frac{1}{4} \left(a_{1}\right)_{1}^{3}+\frac{1}{8} \left(a_{1}\right)_{2}^{2}+\frac{1}{16} \left(a_{1}\right)_{1}^{4}\right) \left(\frac{1}{k} \right)^{3} + \ldots \nonumber \\
\addtocounter{equation}{1}\tag*{\normalsize(\theequation)}
\end{align}}
We can perform a high precision fit of $\frac{\left(a_{0}\right)_{k+1}}{4 \left(a_{0}\right)_{k}} - k^2$ from $k=50$ to $k=100$ and use equation \eqref{e:2DStationaryODE_resurgence?_taylor} to evaluate the coefficients of the series $f_{1}\left(u \right)$, that is $\left(a_{1}\right)_{k}$. In \autoref{table:coefficients obtained from fit and formal transseries solution in 2D} we compare those values with the known $\left(a_{1}\right)_{k}$ of the formal transseries solution. We can see that the prediction of resurgence is valid. Now the only thing left to be determined is the missing multiplying constant $C_1$ in equation \eqref{e:2DStationaryODE_resurgence?}. 
\begin{table}[h!]
\begin{subtable}{1\textwidth}
\centering
\begin{tabular}{|| c | c | c || }
\hline
 & \textbf{\large large $k$ fit} & \textbf{\large transseries } \\
\hline
$\left(a_{1}\right)_{1}$ & -3.1250 & -3.1250 \\
\hline
$\left(a_{1}\right)_{2}$ & 6.4453 & 6.4453 \\
\hline
$\left(a_{1}\right)_{3}$ & -19.1591 & -19.1591 \\
\hline
$\left(a_{1}\right)_{4}$ & 57.7694 & 57.7693 \\
\hline
\end{tabular}
\subcaption{$\lvert B \rvert = 1$} \label{table:2D_B_1_a1ks}
\end{subtable}
\newline
\vspace*{0.1 cm}
\newline
\begin{subtable}{1\textwidth}
\centering
\begin{tabular}{|| c | c | c || }
\hline
 & \textbf{\large large $k$ fit} & \textbf{\large transseries} \\
\hline
$\left(a_{1}\right)_{1}$ & -18.8750 & -18.8750 \\
\hline
$\left(a_{1}\right)_{2}$ & 187.5703 & 187.5703 \\
\hline
$\left(a_{1}\right)_{3}$ & -1536.45022 & -1536.4501 \\
\hline
$\left(a_{1}\right)_{4}$ & 11943.6753 & 11943.6668 \\
\hline
\end{tabular}
\subcaption{$\lvert B \rvert = 2.5$} \label{table:2D_B_2.5_a1ks}
\end{subtable}
\caption{$\left(a_{1}\right)_{k}$ obtained from fit and formal transseries solution in 2D} \label{table:coefficients obtained from fit and formal transseries solution in 2D}
\end{table}
Similarly we check the predictions of resurgence for the 3D case in \autoref{table:coefficients obtained from fit and formal transseries solution in 3D}.
\begin{table}[h!]
\begin{subtable}{1\textwidth}
\centering
\begin{tabular}{|| c | c | c || }
\hline
 & \textbf{\large large $k$ fit} & \textbf{\large transseries} \\
\hline
$\left(a_{1}\right)_{1}$ & 0 & 0 \\
\hline
$\left(a_{1}\right)_{2}$ & 0 & 0 \\
\hline
$\left(a_{1}\right)_{3}$ & -2 & -2 \\
\hline
$\left(a_{1}\right)_{4}$ & 3 & 3 \\
\hline
\end{tabular}
\subcaption{$\lvert B \rvert = 1$} \label{table:3D_B_1_a1ks}
\end{subtable}
\newline
\vspace*{0.1 cm}
\newline
\begin{subtable}{1\textwidth}
\centering
\begin{tabular}{|| c | c | c || }
\hline
 & \textbf{\large large $k$ fit} & \textbf{\large transseries} \\
\hline
$\left(a_{1}\right)_{1}$ & 0 & 0 \\
\hline
$\left(a_{1}\right)_{2}$ & 0 & 0 \\
\hline
$\left(a_{1}\right)_{3}$ & -12.5 & -12.5\\
\hline
$\left(a_{1}\right)_{4}$ & 18.75 & 18.75 \\
\hline
\end{tabular}
\subcaption{$\lvert B \rvert = 2.5$} \label{table:3D_B_2.5_a1ks}
\end{subtable}
\caption{$\left(a_{1}\right)_{k}$ obtained from fit and formal transseries solution in 3D} \label{table:coefficients obtained from fit and formal transseries solution in 3D}
\end{table}
Going back to the 2D case and assuming for a moment that all the coefficients in the series $f_{0}\left(u \right)$ are given by \eqref{e:2DStationaryODE_resurgence?} i.e. $\sum_{k=0}^{\infty} \left(a_{0}\right)_{k} u^{2k}$, writing the gamma function using its integral representation, swapping integration with summation and simplifying gives a series of integrals which give an imaginary contribution 
 \beq
 \Im\left(F_0\left(u \right)\right) \sim C_1 \left(\mp \frac{\pi}{2} \left(- {\mathrm e}^{-\frac{1}{u}} \sqrt{u} \,f_{1}\left(u \right) + \ldots\right)\right)
 \eeq
 where the ``$-$'' sign in front is when the Laplace integral is performed just above the singular direction and the ``$+$'' sign in front is when the Laplace integral is performed just below the singular direction. This ensures that the coefficient (at the lowest order) of $({\sigma_{\Re}})^0$ in equation \eqref{e:Im part of resummed f_0 transseries} i.e. $\Im\left(F_0\left(u \right)\right) + \sigma_{\Im}\, {\mathrm e}^{-\frac{1}{u}} \sqrt{u} \,f_{1}\left(u \right)$, is $0$, when $C_{1} \sim \mp \frac{2}{\pi} \sigma_{\Im}$, and thus satisfying the reality condition on the Laplace-Borel resummation of the formal asymptotic transseries solution of $\rho \left(u \right)$ at the lowest order. This results into a complete expression for resurgence from \eqref{e:2DStationaryODE_resurgence?} as follows
%\begin{equation} 
\begin{align}\label{e:2DStationaryODE_resurgence}
\left(a_{0}\right)_{k} 
\approx & \mp \left\{ -\frac{2}{\pi} \sigma_{\Im} \Gamma  \left(2 k -\frac{1}{2}\right) \left(1+\frac{\left(a_{1}\right)_{1}}{2 k -\frac{3}{2}}+\frac{\left(a_{1}\right)_{2}}{\left(2 k -\frac{3}{2}\right) \left(2 k -\frac{5}{2}\right)} \right. \right. \nonumber \\
& \left.\left. \ \ \ +\frac{\left(a_{1}\right)_{3}}{\left(2 k -\frac{3}{2}\right) \left(2 k -\frac{5}{2}\right) \left(2 k -\frac{7}{2}\right)} + \ldots\right) + \ldots \right\}
\end{align}
%\end{equation}
where, $\left(a_{j}\right)_{k}$ is $k^{th}$ coefficient in a perturbative series around $j^{th}$ exponential term in 2D. We pick the ``$-$'' sign in front of \eqref{e:2DStationaryODE_resurgence} when Laplace-Borel resummation of $f_0\left(u \right)$ is done just above the real axis, and pick the ``$+$'' sign when it is done below.

Using the coefficients $\left(a_{0}\right)_{k}$ and $\left(a_{1}\right)_{k}$ obtained from the formal transseries solution, we can evaluate $\sigma_{\Im}$ from equation \eqref{e:2DStationaryODE_resurgence}. Note that on the left-hand side in equation \eqref{e:2DStationaryODE_resurgence}, the $\left(a_{0}\right)_{k}$ themselves are negative, and therefore $\sigma_{\Im}<0$ when Laplace-Borel resummation is performed just above the real positive line in the Borel plane and $\sigma_{\Im}>0$ when Laplace-Borel resummation is performed just below the real positive line in the Borel plane. We stick to the Laplace-Borel resummation performed just below the real positive line in the Borel plane in the rest of the calculations and that is why we expect $\sigma_{\Im}>0$. Furthermore, only $f_0\left(u \right)$ was Laplace-Borel resummed in equations \eqref{e:Re part of resummed f_0 transseries} and \eqref{e:Im part of resummed f_0 transseries} to understand the resurgence. However, to obtain the actual resummation, we need to Borel sum all $f_j\left(u \right)$ as in \eqref{e:Re part of resummed transseries} and \eqref{e:Im part of resummed transseries}.

\begin{figure}[H]
    \centering
    \begin{subfigure}{0.65\textwidth}
        \centering
        \includegraphics[width=\linewidth]{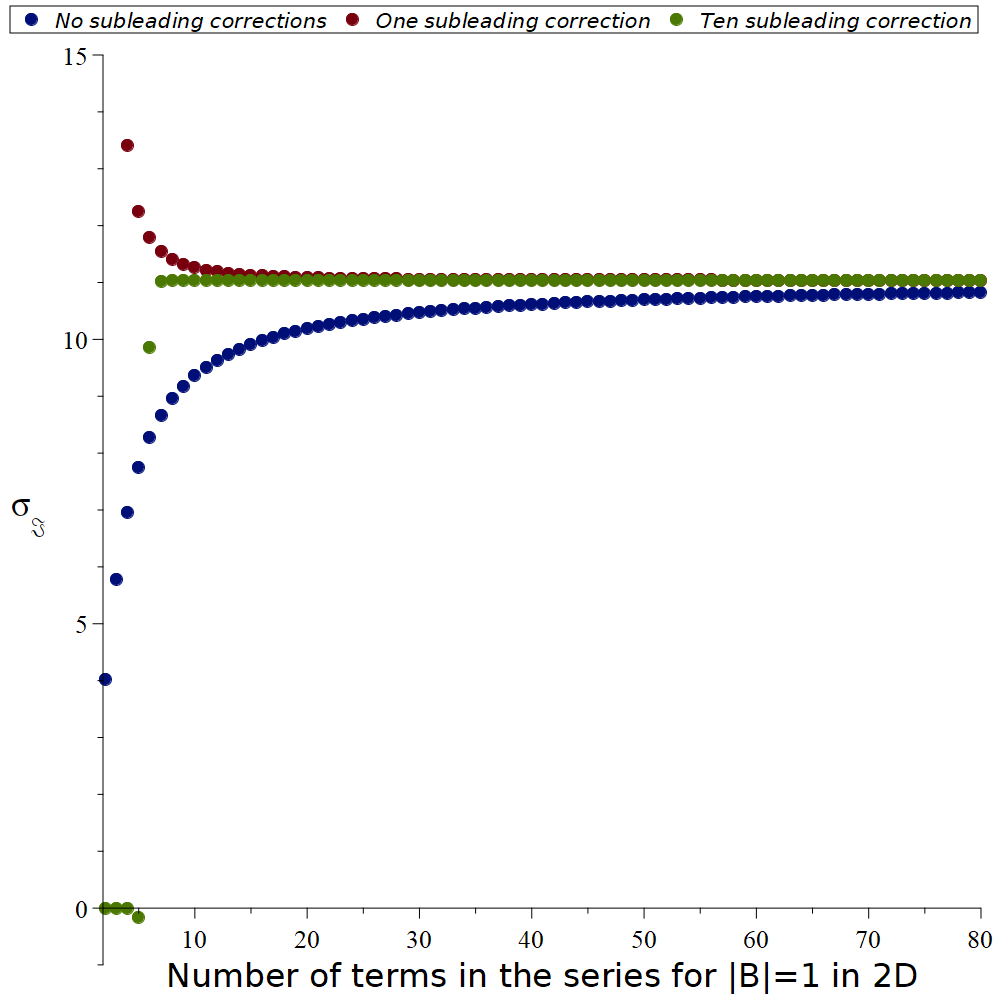}
        \caption{$|B| = 1$} \label{fig:B1_sigmaI_2D}
    \end{subfigure}
    \\[10pt]
    \begin{subfigure}{0.65\textwidth}
        \centering
        \includegraphics[width=\linewidth]{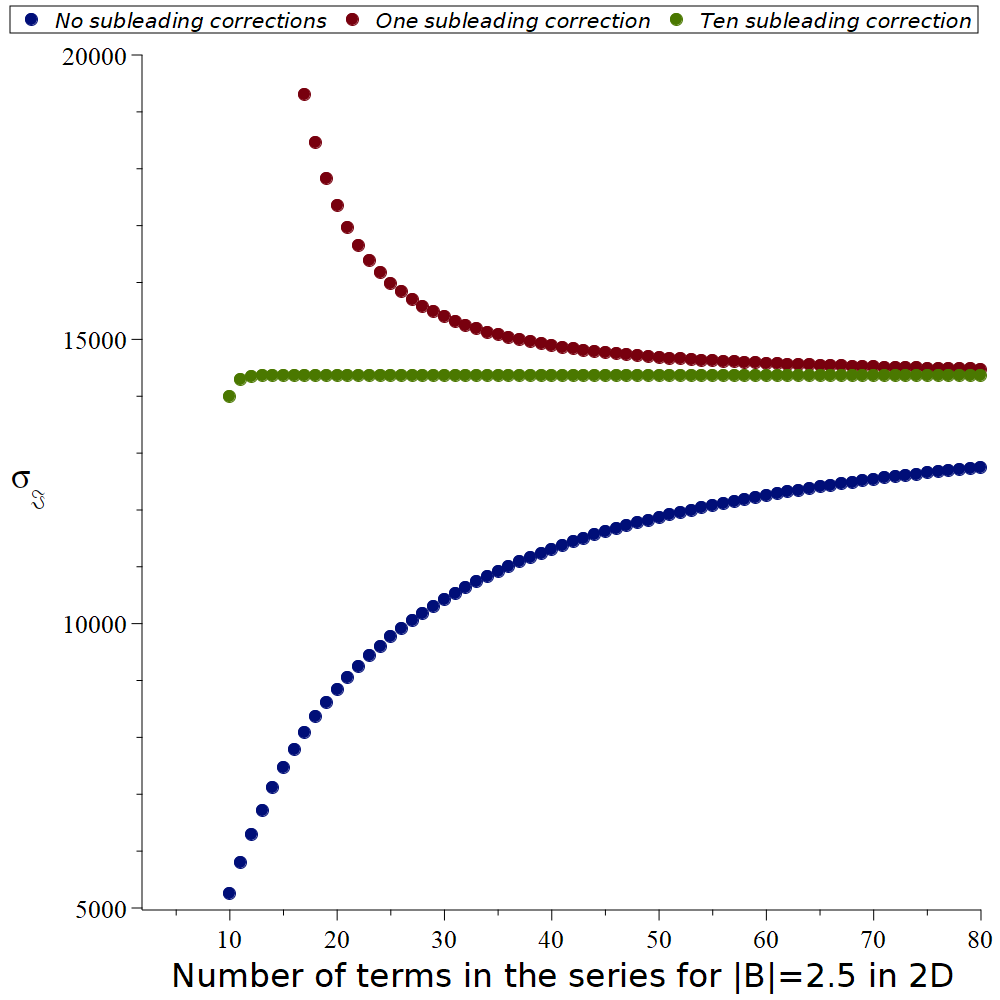}
        \caption{$|B| = 2.5$} \label{fig:B2.5_sigmaI_2D}
    \end{subfigure}
    \vspace{5pt}
    \caption{$\sigma_{\Im}$ vs. number of terms, predicted by resurgence in 2D}
    \label{fig:2D_sigmaIm}
\end{figure}

\begin{figure}[H]
    \centering
    \begin{subfigure}{0.65\textwidth}
        \centering
        \includegraphics[width=\linewidth]{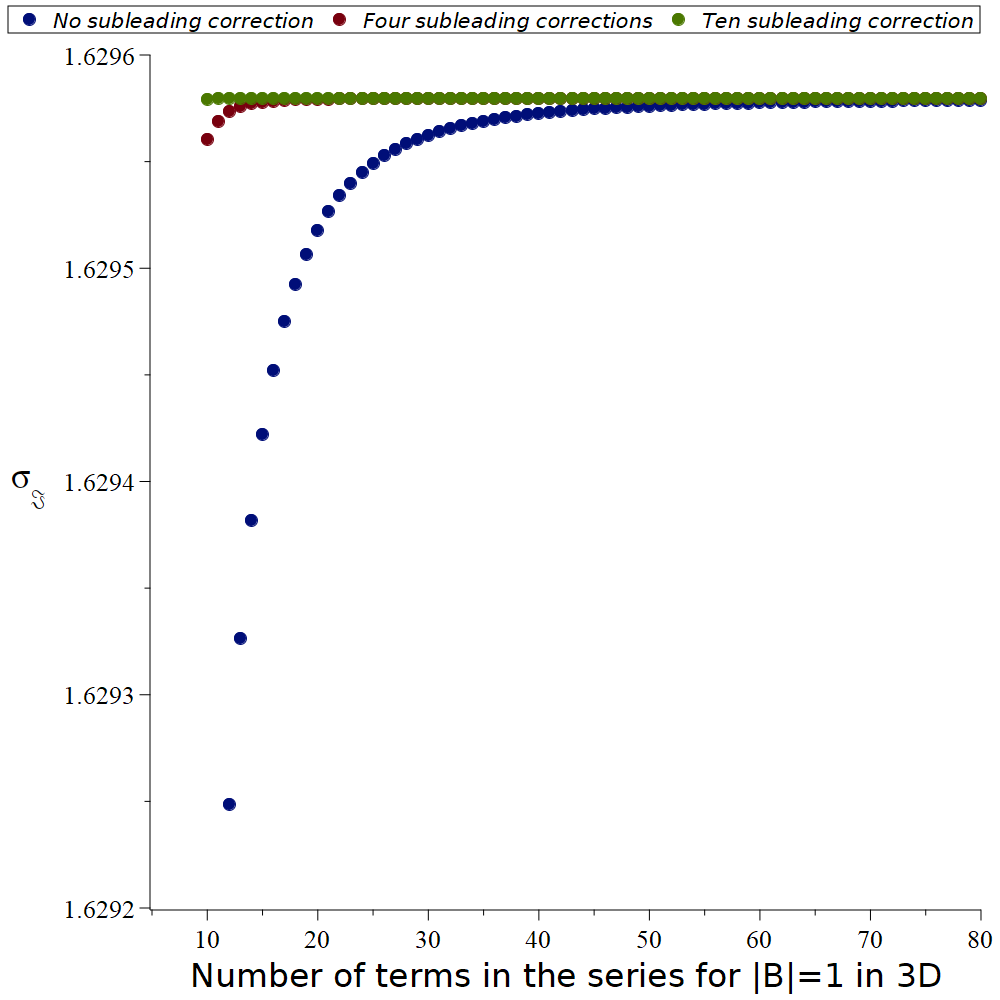}
        \caption{$|B| = 1$} \label{fig:B1_sigmaI_3D}
    \end{subfigure}
    \\[10pt]
    \begin{subfigure}{0.65\textwidth}
        \centering
        \includegraphics[width=\linewidth]{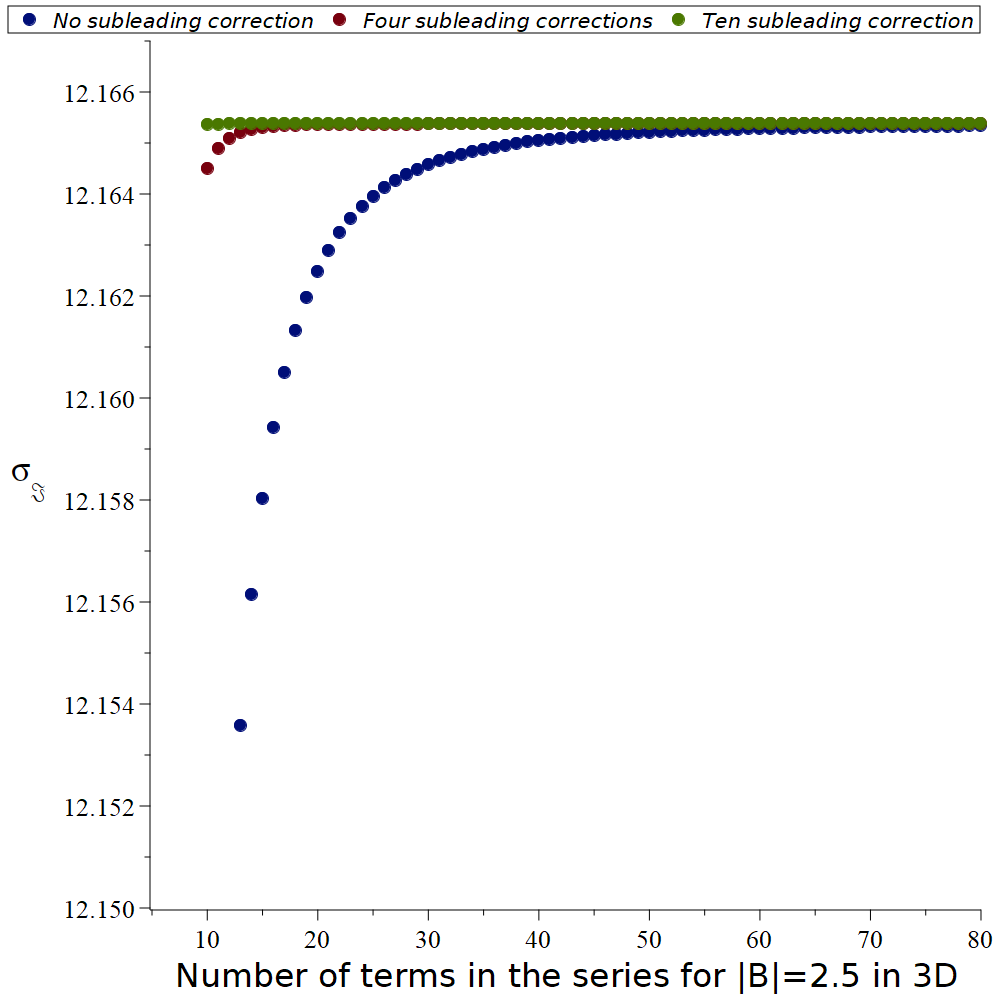}
        \caption{$|B| = 2.5$} \label{fig:B2.5_sigmaI_3D}
    \end{subfigure}
    \vspace{5pt}
    \caption{$\sigma_{\Im}$ vs. number of terms, predicted by resurgence in 3D}
    \label{fig:3D_sigmaIm}
\end{figure}

Given the reality condition on $\rho  \left(u \right)$, $\sigma_{\Im}$ (in 2D) can be obtained numerically from~\eqref{e:2DStationaryODE_resurgence} (considering ``$+$'' sign in front) as discussed earlier, considering only the dominant contribution coming from the part that is linear in $\sigma_{\Im}$ (\cite{Aniceto2015} and \cite{PhysRevD.92.125011}) and with just first few sub-leading corrections. The same procedure can be implemented in 3D $(\mathbf{d}=3)$ by writing a 3D equivalent of equation \eqref{e:2DStationaryODE_resurgence}. The values of $\sigma_{\Im}$ for $|B| = 1$ and $2.5$ in 2D and 3D are shown in Figures \ref{fig:2D_sigmaIm} and \ref{fig:3D_sigmaIm}. Note that although this is a one-parameter transseries with one complex parameter, the imaginary part of the complex parameter is fixed by the reality condition. Therefore, we are left with only one real free parameter $\sigma_{\Re}$, which is equivalent to choosing either the function or its derivative at large $R$ if we want to solve this as an initial value problem. 

As we can see in~\eqref{e:2DStationaryODE_resurgence}, the leading low-order behavior of the coefficients of all the subsequent perturbative sectors shows up in the sub-leading large-order behavior of the coefficients of $f_{0} \left(u \right)$. It can also be shown that the large-order growth of the coefficients of $f_{j} \left(u \right)$ for all $j>0$, has a similar but slightly more complicated resurgence behavior in the sub-leading corrections due to the resurgence of leading low-order coefficients of both the subsequent as well as the previous perturbative sectors.

\subsection{Trans-asymptotic summation}
Another, simpler but less accurate, way to obtain an approximate solution and further check the results, is trans-asymptotic summation (\cite{Costin1999}, \cite{Costin2015}, \cite{PhysRevD.92.125011}). For this we swap the order of summation in power series and summation in exponential non-perturbative terms as follows (in 2D for example) in the transseries ansatz \eqref{e:transseries_ansatz}
\begin{equation} \label{e:Trans-asymptotic series}
\rho  \left(u \right) = \sum_{j=0}^\infty G_{j}(\zeta) u^{j}
\end{equation}
where, $\zeta = \left(\sigma_{\Re} + i \sigma_{\Im} \right) e^{-\frac{1}{u}} \sqrt{u}$.

Substituting \eqref{e:Trans-asymptotic series} into \eqref{e:2DStationaryODEcoordinate_u}, we can write the ODEs for $G_{j}(\zeta)$ in the variable $\zeta$ by grouping the same powers of $u$. These can be solved exactly for up to $j=3$ (see \eqref{e:Trans-asymptotic exact sol}), where the arbitrary constants in $G_{j}(\zeta)$ solutions are fixed by comparing the series expansion of \eqref{e:Trans-asymptotic series} around $\zeta=0$ with the formal transseries solution \eqref{e:transseries_ansatz}. 

{\scriptsize\begin{align} \label{e:Trans-asymptotic exact sol}
G_{0}(\zeta) & = \frac{-\zeta -2}{\zeta -2} \nonumber\\
G_{1}(\zeta) & = \frac{\zeta  \left(-12 B^{2}-\frac{1}{2}\right)}{\left(\zeta -2\right)^{2}}+\frac{\zeta^{2} \left(\zeta -16\right)}{12 \left(\zeta -2\right)^{2}} \nonumber\\
G_{2}(\zeta) & = \frac{\zeta  \left(\frac{55}{2} B^{2}-18 B^{4}-\frac{37}{32}\right)}{\left(\zeta -2\right)^{2}}+\frac{1}{288 \left(\zeta +2\right) \left(\zeta -2\right)^{3}} \Bigg( -\zeta^{6}+\left(-72 B^{2}-41\right) \zeta^{5}+\left(576 B^{2}+588\right) \zeta^{4} \Bigg. \nonumber \\
& \Bigg.+\left(-20736 B^{4}-31680 B^{2}-3492\right) \zeta^{2}+\left(-41472 B^{4}+18432 B^{2}-1656\right) \zeta +4608 B^{2} \Bigg) \nonumber\\
G_{3}(\zeta) & = \frac{\zeta  \left(\frac{16531}{6912}+\frac{371}{4} B^{4}-\frac{2803}{32} B^{2}-18 B^{6}\right)}{\left(\zeta -2\right)^{2}} \nonumber \\ 
& -\frac{216 \zeta}{\left(\zeta +2\right)^{2} \left(\zeta -2\right)^{4}} \Bigg(\frac{4639}{23328}-\frac{\zeta^{8}}{1492992}+\frac{\left(-\frac{65}{72}-B^{2}\right) \zeta^{7}}{10368}+\frac{\left(-\frac{3781}{5184}-B^{4}-\frac{17}{4} B^{2}\right) \zeta^{6}}{576} \Bigg. \nonumber \\ 
& \Bigg.+\frac{\left(B^{4}+\frac{451}{36} B^{2}+\frac{19465}{5184}\right) \zeta^{5}}{144}+\left(B^{6}+\frac{19}{72} B^{4}-\frac{11609}{5184} B^{2}-\frac{29417}{124416}\right) \zeta^{3} \Bigg. \nonumber \\ 
& \Bigg. +\left(4 B^{6}-\frac{115}{36} B^{4}+\frac{77}{27} B^{2}+\frac{4337}{186624}\right) \zeta^{2}+\left(4 B^{6}+\frac{209}{18} B^{4}+\frac{3209}{432} B^{2}+\frac{19889}{31104}\right) \zeta +10 B^{4}-\frac{143 B^{2}}{27} \Bigg) \nonumber\\
\addtocounter{equation}{1}\tag*{\normalsize(\theequation)}
\end{align}}

The real part of trans-asymptotic summation \eqref{e:Trans-asymptotic series} (for $j=0\ldots3$) is in good agreement with the numerical solutions (BVP and BVP+IVP) as shown in Figure \ref{fig:richness of trans-asymptotic1} and the corresponding imaginary parts are $\sim 0$ in the range for which the numerical solutions match well with the real parts of trans-asymptotic summation as shown in \ref{fig:Im richness of trans-asymptotic1}.

\begin{figure}[H] 
\centering
    \includegraphics[width=\linewidth]{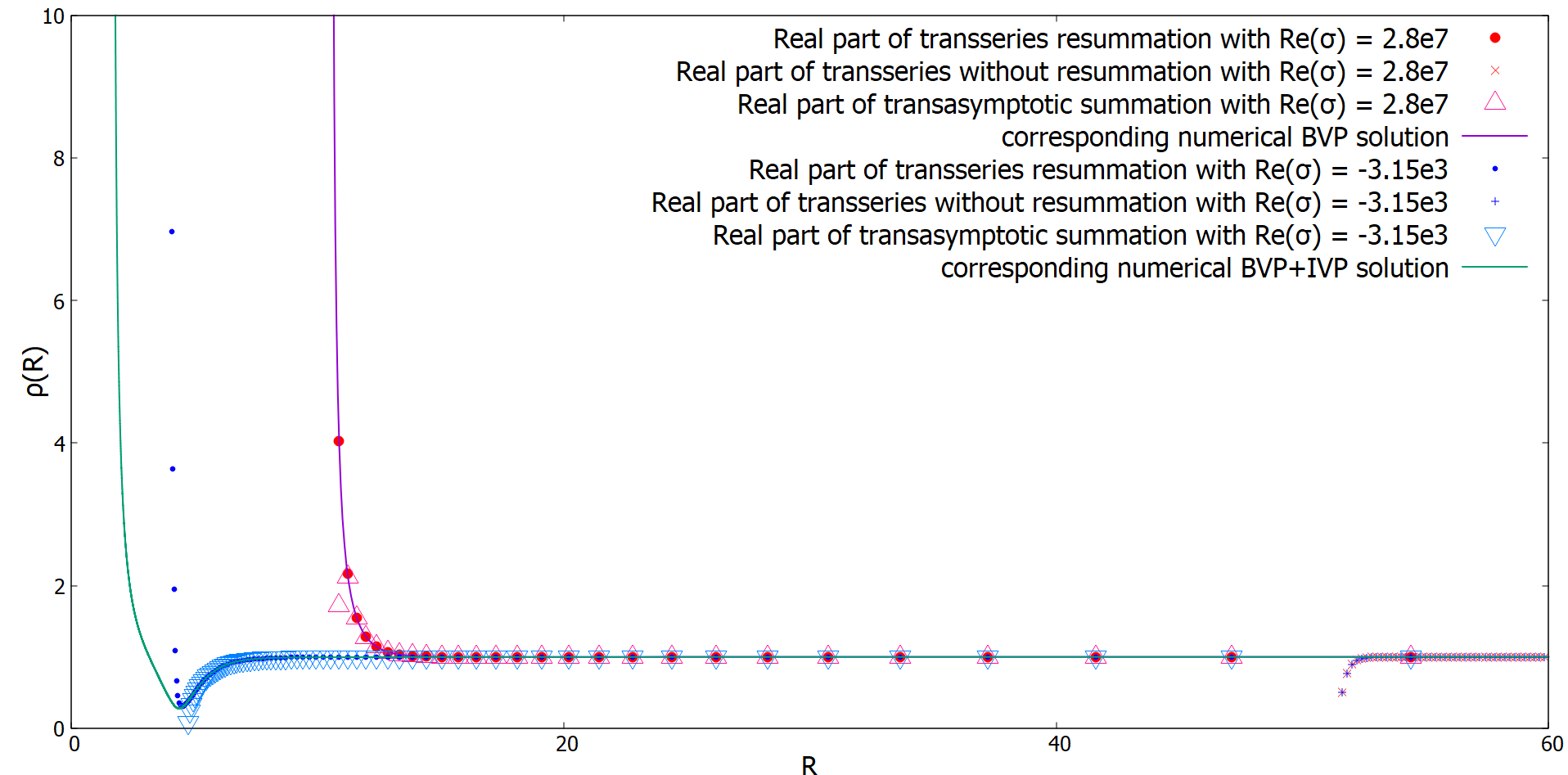}
    \vspace{5pt}
    \caption{Numerical solution, Laplace-Borel resummation, and trans-asymptotic summation} \label{fig:richness of trans-asymptotic1}
\end{figure}

\begin{figure}[H] 
\centering
    \includegraphics[width=\linewidth]{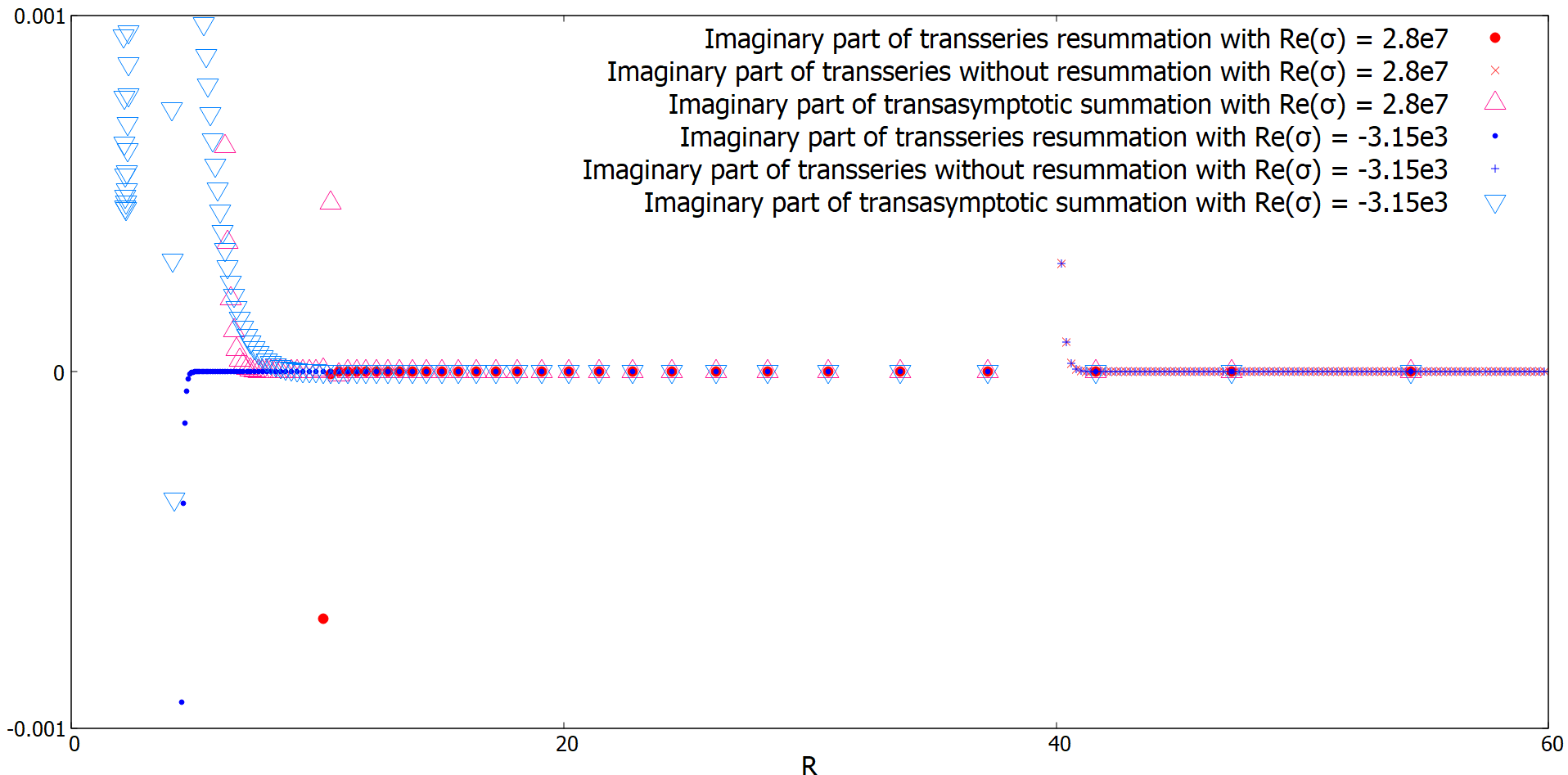}
    \vspace{5pt}
    \caption{Imaginary part of Laplace-Borel resummation and trans-asymptotic summation for the solution \ref{fig:richness of trans-asymptotic1}} \label{fig:Im richness of trans-asymptotic1}
\end{figure}

\iffalse
\begin{figure}[H] 
\centering
    \includegraphics[width=\linewidth]{images_with_better_legends/B1_InvRho_how_good_transasympt_is.png}
    \vspace{5pt}
    \caption{Numerical solution and trans-asymptotic summation} \label{fig:richness of trans-asymptotic1}
\end{figure}

\begin{figure}[H] 
\centering
    \includegraphics[width=\linewidth]{images_with_better_legends/B1_Rho_how_good_transasympt_is_Im.png}
    \vspace{5pt}
    \caption{Imaginary part of trans-asymptotic summation for the solution \ref{fig:richness of trans-asymptotic1} with singularity at $R_{0} \sim 10.5303$} \label{fig:Im richness of trans-asymptotic1}
\end{figure}

\begin{figure}[H] 
\centering
    \includegraphics[width=\linewidth]{images_with_better_legends/B1_InvRho_how_good_transasympt_is_Other_family_of_sol.png}
    \vspace{5pt}
    \caption{A sample solution for $\frac{1}{\rho  \left(R \right)}$ in 2D with $|B| = 1$, obtained using BVP (Boundary Value Problem)+IVP (Initial Value Problem), and matched with the trans-asymptotic summation result} \label{fig:richness of trans-asymptotic2}
\end{figure}

\begin{figure}[H] 
\centering
    \includegraphics[width=\linewidth]{images_with_better_legends/B1_Rho_how_good_transasympt_is_Im_Other_family_of_sol.png}
    \vspace{5pt}
    \caption{Imaginary part of trans-asymptotic summation result for the other family of solution \ref{fig:richness of trans-asymptotic2} with singularity at $R_{0} \sim 1.6643$} \label{fig:Im richness of trans-asymptotic2}
\end{figure}
\fi

\section{Conclusions}

In this paper we considered stationary solutions of the Gross-Pitaevskii equation describing the radial inflow (or outflow) of a superfluid looking for configurations that can be associated with sonic black holes (or white holes). The solutions were found numerically by different procedures and also by transseries expansion and subsequent Borel-Laplace resummation and resurgence. All procedures agree giving a consistent picture of the solutions. We indeed found solutions where the fluid transitions from subsonic to supersonic and therefore can be associated with sonic black holes. We expect that these type of solutions approximately reproduce experimental and theoretical results in \cite{PhysRevA.109.023305} and \cite{Tamura:2023mby}.

\section{Acknowledgments}

We are grateful to the DOE for support under the Fermilab Quantum Consortium. We are grateful to Dr. Sergei Khlebnikov and Dr. Chen-Lung Hung for various comments and discussions. S.V wishes to acknowledge valuable discussions with Akilesh Venkatesh and Kunaal Joshi on numerical techniques. We also wish to acknowledge valuable computational resources provided by RCAC (Rosen Center For Advanced Computing) at Purdue University.

%\nocite{*}
\printbibliography

\appendix
\numberwithin{equation}{section}

\newpage
\section{Resurgence phenomenon in the coefficients of series $f_{1} \left(u \right)$} \label{Appendix A}
{
The coefficients $\left(a_{1}\right)_{k}$ of series $f_{1} \left(u \right)$ in the transseries \eqref{e:transseries_ansatz} for stationary ODE also show a resurgence phenomenon similar to \eqref{e:2DStationaryODE_resurgence}. Naively we might think that the resurgence relation in 2D is of the form
\begin{align} \label{e:2DStationaryODE_resurgence_a1}
\left(a_{1}\right)_{k} 
\approx & \mp \left\{ -\frac{1}{\pi} \sigma_{\Im} \Gamma  \left(k -\frac{1}{2}\right) \left(2 \left(a_{2}\right)_{0}+\frac{2\left(a_{2}\right)_{1}}{k -\frac{3}{2}}+\frac{2 \left(a_{2}\right)_{2}}{\left(k -\frac{3}{2}\right) \left(k -\frac{5}{2}\right)} \right. \right. \nonumber \\
& \left.\left. \ \ \ +\frac{2 \left(a_{2}\right)_{3}}{\left(k -\frac{3}{2}\right) \left(k -\frac{5}{2}\right) \left(k -\frac{7}{2}\right)} + \ldots\right) + \ldots \right\}
\end{align}
so that imaginary ambiguity from \eqref{e:Im part of resummed transseries}, in a similar way as \eqref{e:Im part of resummed f_0 transseries}, is cancelled out in form of $\sigma_{\Re} {\mathrm e}^{-\frac{1}{u}}  \left(\sqrt{u} \Im\left(F_1\left(u \right)\right) +2 \sigma_{\Im}\, {\mathrm e}^{-\frac{1}{u}} u\, f_{2}\left(u \right) \right)$. However, this resurgence relation fails to match the ratio of $\frac{\left(a_{1}\right)_{k+1}}{\left(a_{1}\right)_{k}}$ obtained from the known coefficients of the series $f_{1} \left(u \right)$ for large $k$ values. 

To better understand this, let us consider a two-parameter transseries solution of the form  \cite{Aniceto2015} given below
\begin{equation} \label{e:transseries_ansatz_2parameters}
\rho  \left(u \right) =  \sum_{i=0}^\infty \sum_{j=0}^\infty \sum_{l=0}^{\mathrm{min}(i,j) - j \delta_{ij}} f_{ijl} \left(u \right) {\mathrm e}^{-\frac{\left(i-j\right)}{u}} {u}^{\frac{(\mathbf{d}-1) \left(i-j\right)}{2}} {\left(ln(u) \right)}^{l} {C_{1}}^{i} {C_{2}}^{j}
\end{equation}
where, $f_{n00}$ series are exactly the same as $f_{n}$ series for one-parameter transseries \eqref{e:transseries_ansatz}.

It turns out that resurgence relation of $\left(a_{1}\right)_{k}$ in one-parameter transseries not only contains $\left(a_{2}\right)_{k}$ as in \eqref{e:2DStationaryODE_resurgence_a1}, but also contains $\left(a_{110}\right)_{k}$ \Big(coefficients of the series $f_{110} \left(u \right)$\Big) of the two-parameter transseries \eqref{e:transseries_ansatz_2parameters} (similar to \cite{10.1093/imrn/rnr029}) even though the $f_{110}(u)$ series (which has odd powers of $u$) does not even exist\customfootnote{It can be easily shown that these $\left(a_{110}\right)_{k}$s are complex nonlinear functions of $\left(a_{000}\right)_{k}$s, $\left(a_{010}\right)_{k}$s and $\left(a_{100}\right)_{k}$s. Furthermore, $\left(a_{010}\right)_{k}$ are also complex nonlinear functions of $\left(a_{100}\right)_{k}$s and $\left(a_{000}\right)_{k}$s. Additionally, $\left(a_{n00}\right)_{k}$s of two-parameter transseries are just $\left(a_{n}\right)_{k}$s of one-parameter transseries. This means that effectively we are still using the coefficients of one parameter transseries even in this resurgence relation.} in the one-parameter transseries \eqref{e:transseries_ansatz}.

The actual resurgence relation, from \eqref{e:2DStationaryODE_resurgence_a1} becomes
{\footnotesize\begin{align} \label{e:2DStationaryODE_resurgence_a1_actual}
\left(a_{1}\right)_{k} 
\approx & \mp \left\{ -\frac{1}{\pi} \sigma_{\Im} \Gamma  \left(k -\frac{1}{2}\right) \left(2 \left(a_{2}\right)_{0} + {\left(-1\right)}^{k} \left(a_{110}\right)_{0} +\frac{2\left(a_{2}\right)_{1}}{k -\frac{3}{2}}+\frac{2 \left(a_{2}\right)_{2} + {\left(-1\right)}^{k} \left(a_{110}\right)_{1}}{\left(k -\frac{3}{2}\right) \left(k -\frac{5}{2}\right)} \right. \right. \nonumber \\
& \left.\left. \ \ \ +\frac{2 \left(a_{2}\right)_{3}}{\left(k -\frac{3}{2}\right) \left(k -\frac{5}{2}\right) \left(k -\frac{7}{2}\right)} +\frac{2 \left(a_{2}\right)_{4} + {\left(-1\right)}^{k} \left(a_{110}\right)_{2}}{\left(k -\frac{3}{2}\right) \left(k -\frac{5}{2}\right) \left(k -\frac{7}{2}\right) \left(k -\frac{9}{2}\right)} + \ldots\right) + \ldots \right\}
\addtocounter{equation}{1}\tag*{\normalsize(\theequation)}
\end{align}}
where, we pick the ``$+$'' sign in front since we perform the Laplace integration just below the real positive line. The terms with ${\left(-1\right)}^{k}$ do not give an imaginary contribution to the Laplace-Borel resummation. Therefore, the existence of $\left(a_{110}\right)_{k}$s in the resurgence relation \eqref{e:2DStationaryODE_resurgence_a1_actual}, does not really affect the reality condition on the transseries resummation, since the resurgence relation is a tool only for predicting (from the asymptotic behavior) imaginary ambiguity in the Laplace-Borel resummation of a divergent series. The resurgence relation \eqref{e:2DStationaryODE_resurgence_a1_actual} is in 2D, but a similar resurgence relation can also be written in 3D.

Using resurgence relations for $\left(a_{1}\right)_{k}$ in 2D and 3D, with only the first 5 terms, we can get a really good agreement between the ratio $\frac{\left(a_{1}\right)_{k+1}}{\left(a_{1}\right)_{k}}$ obtained from the series coefficients of $f_{1}(u)$ and the same ratio obtained from the right-hand side of the resurgence relations as shown in Figs. \ref{fig:2D a[1][k+1]/a[1][k]} and \ref{fig:3D a[1][k+1]/a[1][k]}.

\begin{figure}[H]
    \centering
    \begin{subfigure}{0.65\textwidth}
        \centering
        \includegraphics[width=\linewidth]{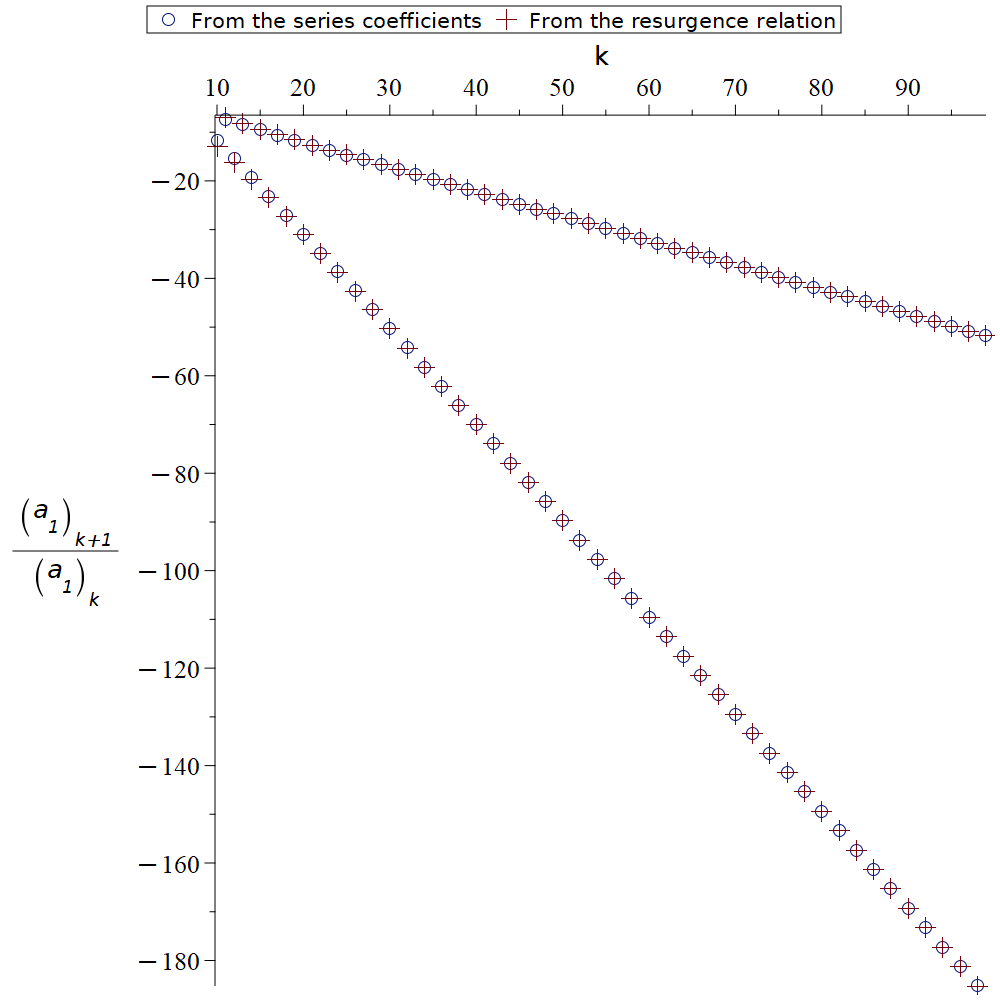}
        \caption{$|B| = 1$} \label{fig:B1_a1_coefficient_growth}
    \end{subfigure}
    \\[10pt]
    \begin{subfigure}{0.65\textwidth}
        \centering
        \includegraphics[width=\linewidth]{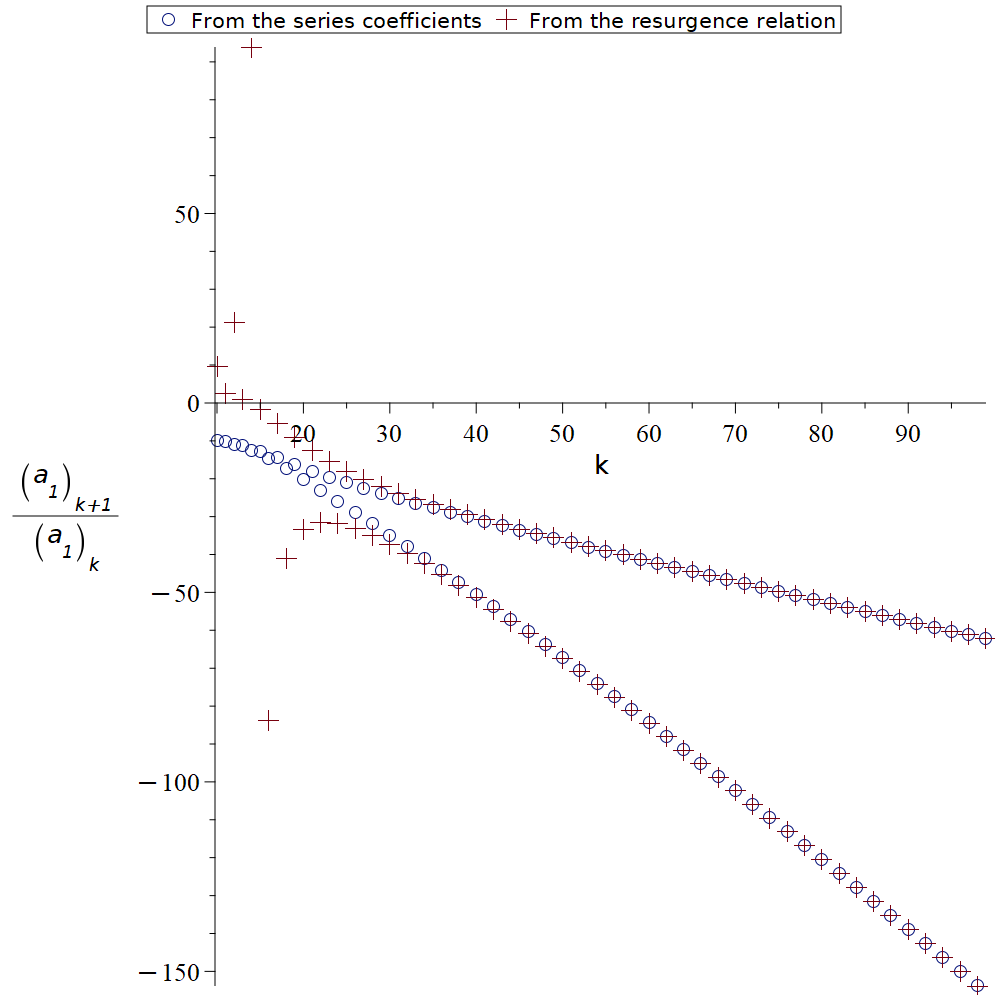}
        \caption{$|B| = 2.5$} \label{fig:B2.5_a1_coefficient_growth}
    \end{subfigure}
    \vspace{5pt}
    \caption{$\frac{\left(a_{1}\right)_{k+1}}{\left(a_{1}\right)_{k}}$ vs. number of terms in 2D}
    \label{fig:2D a[1][k+1]/a[1][k]}
\end{figure}

\begin{figure}[H]
    \centering
    \begin{subfigure}{0.65\textwidth}
        \centering
        \includegraphics[width=\linewidth]{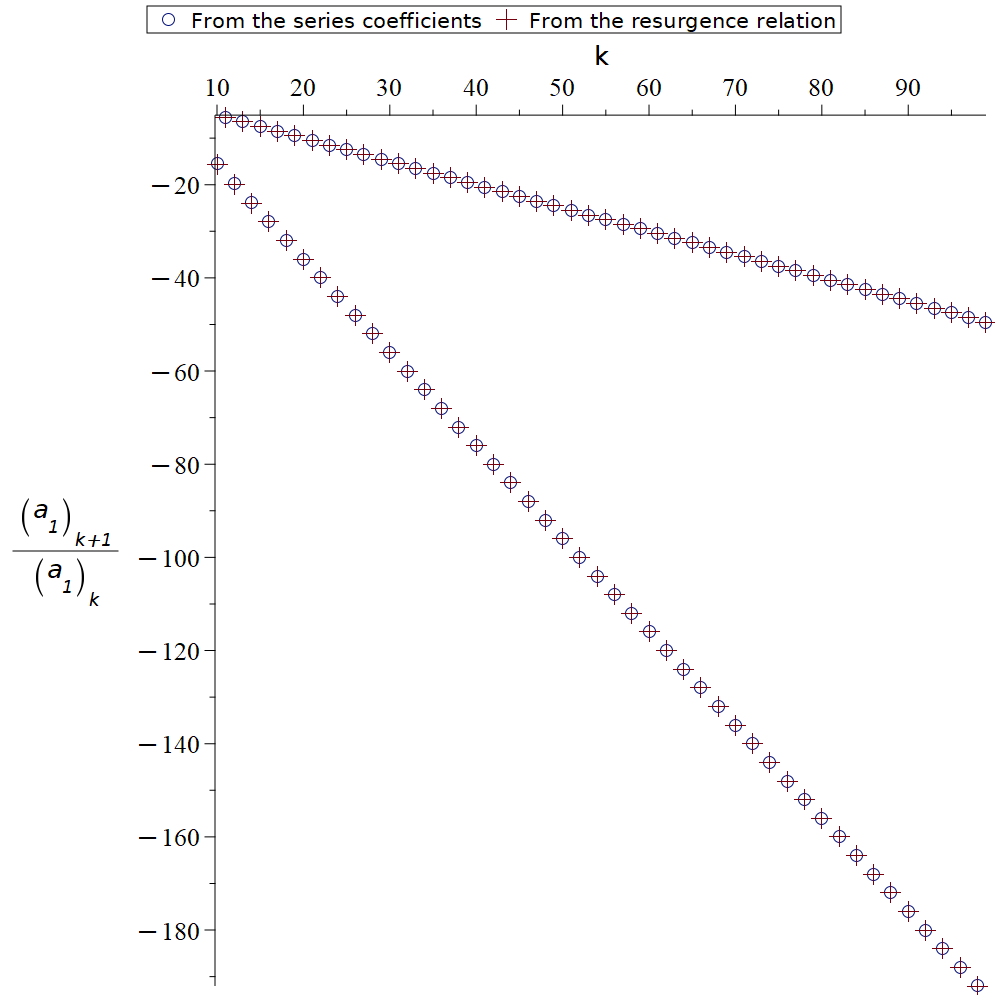}
        \caption{$|B| = 1$} \label{fig:B1_a1_coefficient_growth_3D}
    \end{subfigure}
    \\[10pt]
    \begin{subfigure}{0.65\textwidth}
        \centering
        \includegraphics[width=\linewidth]{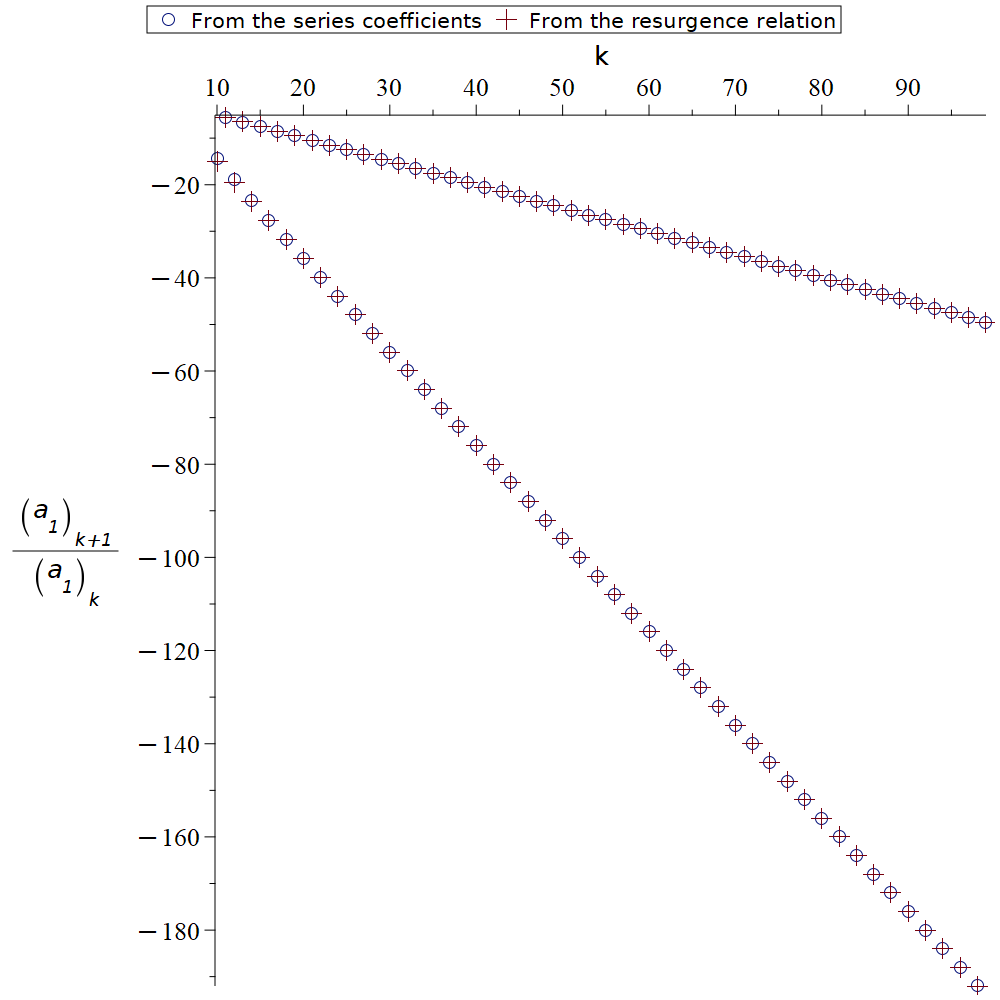}
        \caption{$|B| = 2.5$} \label{fig:B2.5_a1_coefficient_growth_3D}
    \end{subfigure}
    \vspace{5pt}
    \caption{$\frac{\left(a_{1}\right)_{k+1}}{\left(a_{1}\right)_{k}}$ vs. number of terms in 3D}
    \label{fig:3D a[1][k+1]/a[1][k]}
\end{figure}
}

\section{Stationary singular solutions for $\rho(R)$ that go to $0$ as $R\to\infty$} \label{Appendix B}
{
\iffalse
\begin{figure}[H]
    \centering
    \begin{subfigure}{1\textwidth}
        \centering
        \includegraphics[width=\linewidth]{images_with_better_legends/B_1_oscil_3D.png}
        \caption{Sample solution for B=1.0 in 3D} \label{fig:B_1.0_oscil_3D}
    \end{subfigure}
    \\[10pt]
    \begin{subfigure}{1\textwidth}
        \centering
        \includegraphics[width=\linewidth]{images_with_better_legends/B_1_oscil_3D_crossing.png}
        \caption{White hole configuration for \ref{fig:B_1.0_oscil_3D}} \label{fig:B_1.0_oscil_3D_cross}
    \end{subfigure}
    \vspace{5pt}
    \caption{Sample oscillatory stationary solutions}
    \label{fig:Sample oscillatory stationary solutions}
\end{figure}
\fi

\begin{figure}[H]
    \centering
    \includegraphics[width=\linewidth]{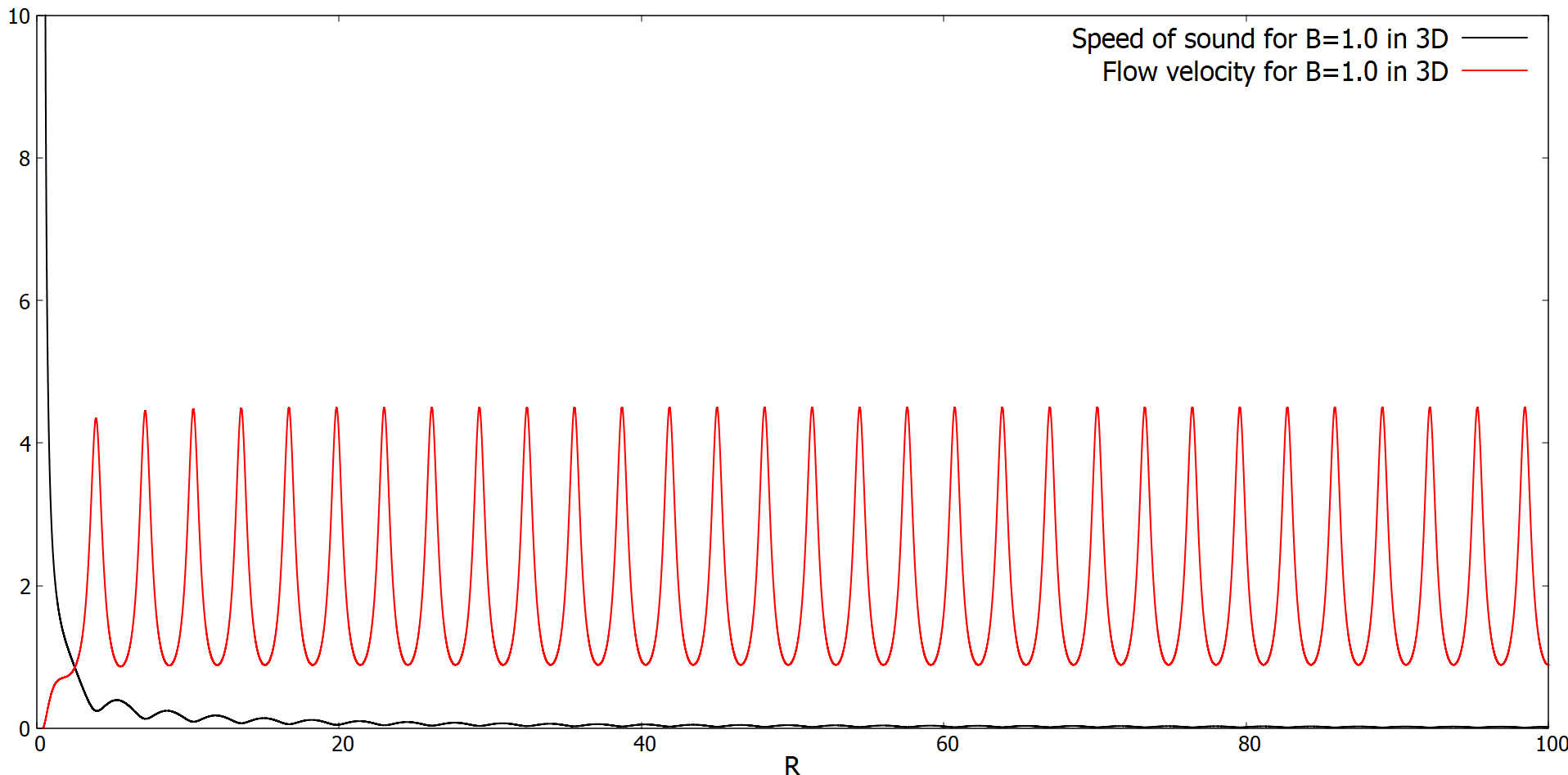}
    \vspace{5pt}
    \caption{White hole configuration for for $B=1$ in 3D} \label{fig:B_1.0_oscil_3D_cross}
\end{figure}

In this paper we focused only on stationary solutions of \eqref{e:StationaryODE} that approach $1$ as $R\to\infty$. However, in principle we can also check for the solutions which go like $\frac{\sqrt{B}}{R^\frac{\mathbf{d}-1}{2}}$ as $R\to\infty$. For such solutions, the flow can be only outward (from \eqref{e:scaled_vel} and \eqref{e:dtheta_dR}) as the $B$ parameter has to be positive for the solutions to be real. If we substitute $\rho  \left(R \right) = \frac{\sqrt{B}(1+f(R)}{R^\frac{\mathbf{d}-1}{2}}$ in~\eqref{e:StationaryODE} and linearize in $f(R)$ \cite{Costin2015}, we get oscillatory non-perturbative corrections of the form $\exp\left(\pm i 2 R\right)$ which are finite everywhere. That means such solutions have a two-parameter asymptotic transseries and none of the parameters has to be $0$. Therefore, this problem can be solved (in 2D and 3D) as an initial value problem by picking some values of $\rho  \left(R \right)$ and $\rho'  \left(R \right)$, starting from some large $R$ and using a method like RK4.\customfootnote{Solutions of the form $\rho\left(R\to\infty\right) \rightarrow \frac{\sqrt{B}}{R^\frac{\mathbf{d}-1}{2}}$ are mentioned only briefly to show how they differ from the solutions of the form $\rho\left(R\to\infty\right) \rightarrow 1$ which are the main focus of this paper.} From the speed of sound \eqref{e:scaled_sound_speed} and the flow velocity (\eqref{e:scaled_vel} and \eqref{e:dtheta_dR}), we see that such solutions have a white hole configuration (with outward flow) as shown for a sample solution in Figure \ref{fig:B_1.0_oscil_3D_cross}.
}

\end{document}